\documentclass[twocolumn]{aastex631}

\usepackage{graphicx}
\usepackage{amsmath}
\usepackage{float}
\usepackage{natbib}
\usepackage{tabularx}
\usepackage{bm}
\usepackage{appendix}
\usepackage{longtable}
\usepackage{enumitem}

\newcommand\msun{M_{\odot}}

\shorttitle{Single MSPs in Globular Clusters}
\shortauthors{Ye et al.}

\graphicspath{{./}{figures/}}

\begin{document}

\title{Single Millisecond Pulsars from Dynamical Interaction Processes in Dense Star Clusters}

\author[0000-0001-9582-881X]{Claire S. Ye}
\affil{Canadian Institute for Theoretical Astrophysics, University of Toronto, 60 St. George Street, Toronto, Ontario M5S 3H8, Canada}
\correspondingauthor{Claire S.~Ye}
\email{claireshiye@cita.utoronto.ca}

\author[0000-0002-4086-3180]{Kyle Kremer}
\altaffiliation{NASA Einstein Fellow}
\affiliation{TAPIR, California Institute of Technology, Pasadena, CA 91125, USA}

\author[0000-0001-5799-9714]{Scott M.\ Ransom}
\affiliation{NRAO, 520 Edgemont Road, Charlottesville, VA 22903, USA}

\author[0000-0002-7132-418X]{Frederic A.\ Rasio}
\affil{Department of Physics \& Astronomy, Northwestern University, Evanston, IL 60208, USA}
\affil{Center for Interdisciplinary Exploration \& Research in Astrophysics (CIERA), Northwestern University, Evanston, IL 60208, USA}

\begin{abstract}
Globular clusters (GCs) are particularly efficient at forming millisecond pulsars. Among these pulsars, about half lack a companion star, a significantly higher fraction than in the Galactic field. This fraction increases further in some of the densest GCs, especially those that have undergone core collapse, suggesting that dynamical interaction processes play a key role. For the first time, we create $N$-body models that reproduce the ratio of single-to-binary pulsars in Milky-Way-like GCs. We focus especially on NGC~6752, a typical core-collapsed cluster with many observed millisecond pulsars. Previous studies suggested that an increased rate of neutron star binary disruption in the densest clusters could explain the overabundance of single pulsars in these systems. Here, we demonstrate that binary disruption is ineffective and instead we propose that two additional dynamical processes play the dominant role: (1) tidal disruption of main-sequence stars by neutron stars; and (2) gravitational collapse of heavy white-dwarf-binary merger remnants. Neutron stars formed through these processes may also be associated with fast radio bursts similar to those observed recently in an extragalactic GC.
\end{abstract}

%\keywords{Classical Novae (251) --- Ultraviolet astronomy(1736) --- History of astronomy(1868) --- Interdisciplinary astronomy(804)}

\section{Introduction} \label{sec:intro_isomsp}

Pulsars are highly magnetized, fast-spinning neutron stars (NSs) ubiquitous in the Galaxy. There are $\sim3400$~pulsars detected in the Milky Way so far \citep[][]{Manchester+2005}\footnote{ATNF Catalog: https://www.atnf.csiro.au/research/pulsar/psrcat/}. Among these pulsars, about 570 have millisecond spin periods ($P\lesssim30$~ms) and low spin-down rates ($\dot P \lesssim 10^{-19}$; \citealp{Lorimer_2008}), thus are called millisecond pulsars (MSPs). They are proposed to be ``recycled" in low-mass X-ray binaries, where an NS accreted material and angular momentum from the companion star and was revived and spun-up to millisecond periods \citep{Alpar+1982}. Given how MSPs were formed, it was unexpected when observations showed that a significant number of MSPs do not have a companion star. Specifically, $\sim20\%$ of the MSPs detected in the Galactic field\footnote{http://astro.phys.wvu.edu/GalacticMSPs/} and $\sim40\%$ of the MSPs detected in globular clusters (GCs)\footnote{GC Pulsar Catalog: \newline https://www3.mpifr-bonn.mpg.de/staff/pfreire/GCpsr.html} are isolated. The large number of single MSPs has motivated previous studies to search for their origins, and a few scenarios have been suggested. These include the ``failed'' double NS binary formation where the binary with an MSP is disrupted by the second supernova explosion \citep{Camilo+1993,Belczynski+2010}, the ``evaporation'' of the low-mass hot white dwarf (WD) companions \citep{Bildsten_2002}, and the merger products of double WDs \citep[e.g.,][]{Schwab_2021}.

\begin{figure}
\begin{center}
\includegraphics[width=\columnwidth]{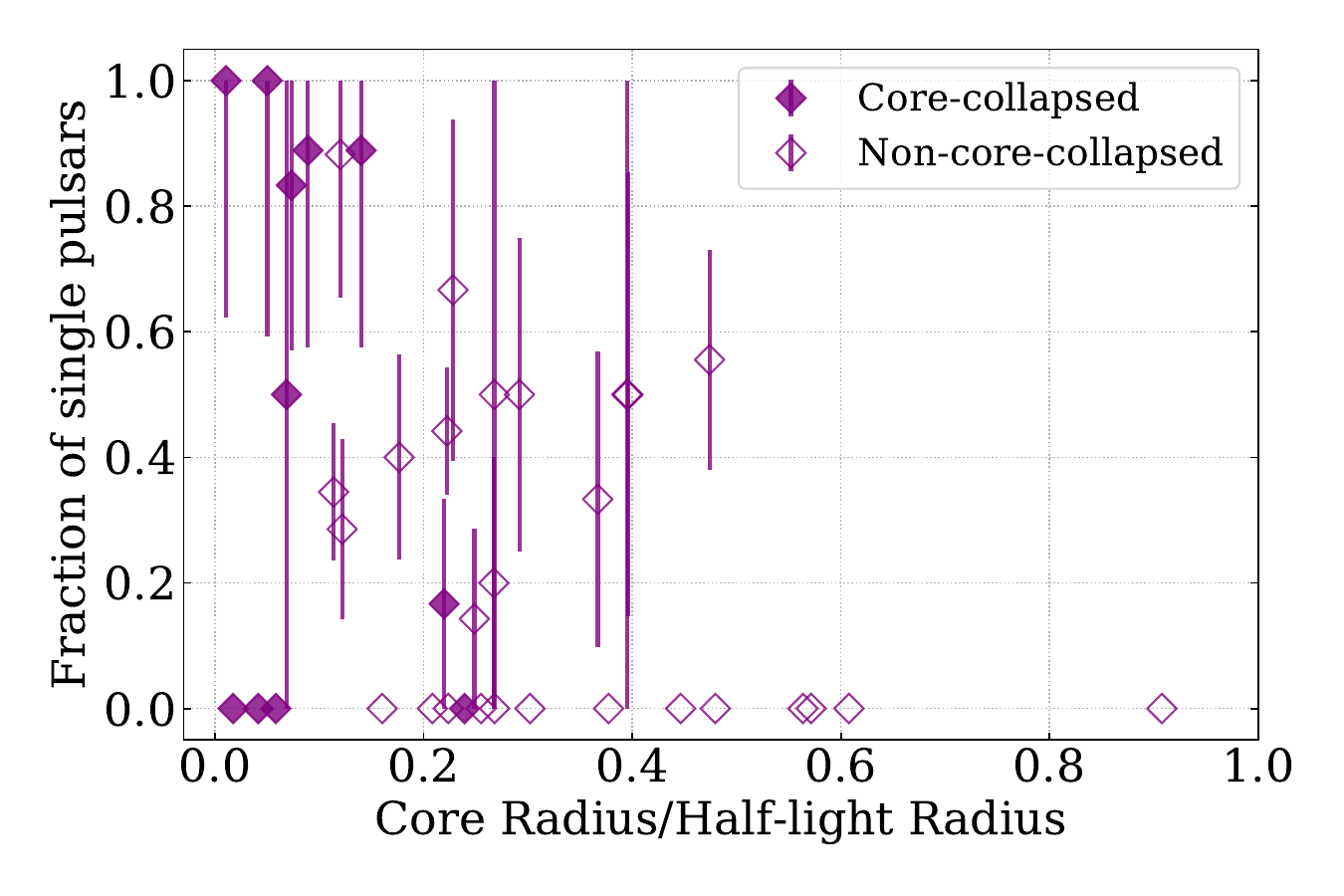}
\caption{Fraction of single pulsars as a function of the ratio of the core-to-half light radii. The diamonds show the observations from the GC Pulsar Catalog. Filled diamonds mark core-collapsed clusters from \citet[][2010 edition]{Harris_1996} and vice versa. The lengths of the error bars reflect the number of observed single pulsars in the clusters. The error bars are calculated as $\sqrt{N_{\rm{sPSR}}}$/$N_{\rm{tPSR}}$, where $N_{\rm{sPSR}}$ and $N_{\rm{tPSR}}$ are the observed number of single and all pulsars in a cluster, respectively. Both the Pearson and Spearman statistical tests show that the fraction of single pulsars is correlated with the core-to-half-light radius with $>2\sigma$ significance and a correlation coefficient of about $-0.4$. The averages of the fractions for core-to-half light radii in the ranges of 0-0.2, 0.2-0.4, and $>0.4$ are 0.50, 0.23, and 0.08. respectively. There are about $15$ observed cluster pulsars with spin periods larger than the typical MSP spin periods, where the maximum observed spin period is about $1000$~ms (there is an observed pulsar with a spin period of $~2500$~ms, but it is not clearly associated with the potential host cluster).}\label{fig:fsmap_compare}
\end{center}
\end{figure}

The relatively high fraction of single MSPs in the GCs indicates an additional contribution from dynamical formation channels. The role of dynamics is further motivated by the fact that single MSPs are most abundant in the dense core-collapsed clusters, where they contribute to $\sim55\%$ of the detected MSPs (Figure~\ref{fig:fsmap_compare} and Table~\ref{tab:obs_npsr}). For example, previous studies have suggested that the disruption of MSP binaries by dynamical interactions \citep{Verbunt_Freire_2014} and NS-main sequence star tidal disruption events (TDEs; \citealp[e.g.,][]{Davies+1992,Davies_Benz_1995,Lee+1996,Camilo_Rasio_2005,Kremer_nstde_2022}) can produce single MSPs.

However, the exact ratios of single-to-binary pulsars in Milky Way GCs (taking into account the aforementioned mechanisms) have yet to be self-consistently reproduced by $N$-body cluster simulations. As an illustration of this, we searched for single MSPs in the \texttt{CMC} Cluster Catalog Models. These catalog models have shown promising results in matching various properties of observed Milky Way GCs including the cluster masses, core radii, and half-light radii, and are close representatives of the Milky Way GCs \citep{Kremer+2020catalog}. The models also form and evolve MSPs self-consistently in the dynamical environment of clusters, producing good agreements with the properties of the observed cluster pulsars \citep{Ye_msp_2019}. However, the fractions of single MSPs in these models overlap with the observations but do not show any difference between core-collapsed and non-core-collapsed clusters as the observations do in Figure~\ref{fig:fsmap_compare}. In addition, the total fraction of single MSPs in all core-collapsed models is $\sim 20\%$, much lower than the $\sim 55\%$ for the observed core-collapsed clusters.

\startlongtable
\begin{deluxetable}{ccccc}
\tabletypesize{\scriptsize}
%\tablewidth{4pt}
%\setlength{\tabcolsep}{3pt}
\tablecaption{Number of Observed Pulsars in GCs\label{tab:obs_npsr}} 
\tablehead{\colhead{Core-collapsed GCs} & \colhead{$r_c$/$r_{hl}$} & \colhead{$N_{\rm sPSR}$} & \colhead{$N_{\rm bPSR}$} & \colhead{$N_{\rm sPSR}$/$N_{\rm tPSR}$}
}
\startdata
Terzan 1 & 0.01 & 7 & 0 & 1.00\\
NGC 6522 & 0.05 & 6 & 0 & 1.00\\
NGC 6752 & 0.09 & 8 & 1 & 0.89\\
NGC 7078 & 0.14 & 8 & 1 & 0.89\\
NGC 6624 & 0.07 & 10 & 2 & 0.83\\
NGC 6342 & 0.07 & 1 & 1 & 0.50\\
NGC 362 & 0.22 & 1 & 5 & 0.17\\
NGC 6397 & 0.02 & 0 & 2 & 0.00\\
NGC 6544 & 0.04 & 0 & 3 & 0.00\\
NGC 7099 & 0.06 & 0 & 2 & 0.00\\
NGC 6266 & 0.24 & 0 & 9 & 0.00\\
\hline
- & 0-0.2 & 75 & 57 & 0.57\\ 
- & 0.2-0.4 & 38 & 79 & 0.32\\
- & $>0.4$ & 10 & 19 & 0.34

\enddata
\tablecomments{For the 11 core-collapsed GCs from \citet[][2010 edition]{Harris_1996} with observed pulsars from left to right: Cluster name, the ratio of core-to-half light radii from \citet[2010 edition]{Harris_1996}, the number of single pulsars, the number of binary pulsars, and the fraction of single pulsars (also shown in Figure~\ref{fig:fsmap_compare} as filled diamonds). NGC~6266 is marked as core-collapsed in \citet[][2010 edition]{Harris_1996} but has no single MSPs. This may be because the cluster has a relatively large core-to-half light ratio compared to other core-collapsed clusters, and thus the dynamical evolution of MSPs in this cluster is probably more similar to non-core-collapsed clusters. Indeed, \citet{Beccari+2006} suggested that this cluster has not undergone core collapse. A low fraction of single MSPs is also observed in NGC~362, which has a similar core-to-half light ratio as NGC~6266. The last three rows show all clusters' numbers and fractions (including non-core-collapsed GCs) for core-to-half light radii in the range of 0-0.2, 0.2-0.4, and $>0.4$ from top to bottom. Denser clusters with smaller core-to-half light radii have higher fractions of single pulsars.}
\end{deluxetable}

The high fraction of observed single MSPs and the discrepancy between the cluster models and observations inspire us to understand quantitatively the dynamical channels for forming single MSPs in GCs. In this study, we use the ratios of single-to-binary MSPs in GCs as guidance for forming MSPs since the ratios are likely less affected by the selection biases of MSPs (e.g., beaming fractions; \citealp{Lorimer_2008}) than the observed number of MSPs. In the following Section~\ref{sec:method}, we will introduce the Monte Carlo $N$-body dynamics code used for self-consistently simulating GCs and the compact object dynamics within, focusing on one cluster NGC~6752. In Section~\ref{sec:pathways}, we show the number of single MSPs formed through different channels. We discuss the uncertainties of these channels in Section~\ref{sec:discuss} and conclude in Section~\ref{sec:conclu}.

\newpage
\section{$N$-body Simulation Methods}\label{sec:method}
\subsection{Cluster Monte Carlo Code}\label{subsec:code}
We use the \texttt{Cluster Monte Carlo} code (\texttt{CMC}; \citealp[][and references therein]{CMC1}) to model GCs and all objects within. \texttt{CMC} is a H\'enon-style Monte Carlo code \citep{henon1971monte,henon1971montecluster} that incorporates various relevant physics for cluster evolution, including two-body relaxation, tidal mass loss, strong few-body interactions, dynamical binary formation through collisions with giant stars and tidal capture interactions \citep{Ye_47tuc_2021}, and post-Newtonian effects \citep{Rodriguez+repeated2018}. Stellar and binary star evolution are fully coupled to the dynamical interactions and are computed by the publicly available software \texttt{COSMIC} \citep{cosmic}, which is based on \texttt{SSE} \citep{hurley2000comprehensive} and \texttt{BSE} \citep{hurley2002evolution}. Strong three- and four-body gravitational encounters are directly integrated using the \texttt{Fewbody} package \citep{fregeau2004stellar,fregeau2007monte}, which includes post-Newtonian dynamics for black holes \citep[BHs;][]{antognini2014rapid,amaro2016relativistic,Rodriguez+repeated2018,rodriguez2018postb}.

\texttt{CMC} forms and evolves NSs and MSPs following \citet{Ye_msp_2019} and references therein. NSs are formed either in iron core-collapse supernovae (CCSNe) or electron-capture supernovae (ECSNe) depending on the mass and metallicity of the progenitor stars. NSs formed in CCSNe receive large natal kicks drawn from a Maxwellian distribution with a standard deviation $\sigma_{\rm{CCSN}}=265\,\rm{km\,s^{-1}}$ \citep{Hobbs+2005}. ECSNe include scenarios where the core of a $\sim6-8\,\msun$ star collapses to an NS \citep{nomoto1984evolution,nomoto1987evolution}, or the collapse of a massive WD to an NS triggered by the capture of electrons and the loss of the electron pressure after mass accretion from or a merger with another WD \citep{nomoto1991conditions,Saio_Nomoto_1985,Saio_Nomoto_1998,Saio_Nomoto_2004,Shen+2012,Schwab+2015,Schwab_2021}. The critical mass for the collapse of an ONe core or WD is assumed to be $1.38\,\msun$. For more details about different outcomes of WD--WD coalescences in \texttt{CMC}, see also \citet{Kremer+2021WD}. NSs formed in ECSNe receive small natal kicks drawn from a Maxwellian distribution with a standard deviation $\sigma_{\rm{ECSN}}=20\,\rm{km\,s^{-1}}$ \citep{Kiel+2008,Kiel_Hurley_2009}. The small kicks allow these NSs to be preferentially retained in the clusters given the low cluster escape velocities. We set the maximum mass of an NS to be $2.5\,\msun$. Any object more massive than this will collapse into a BH.

All new NSs born in CCSNe and ECSNe are assigned initial magnetic fields in the range of $10^{11.5}-10^{13.8}$~G and spin periods in the range of $30-1000$~ms, similar to the observed young radio pulsars (ATNF Catalog). Isolated young pulsars spin down through magnetic dipole radiation, and their surface magnetic fields decay exponentially in a $3$~Gyr timescale.\footnote{Note that the evolution of NS magnetic fields is still under debate, and the timescale for magnetic field decay (if at all) is uncertain \citep[e.g.,][and references therein]{Igoshev+2021}. The decay timescale adopted here can produce the observed cluster pulsars reasonably well \citep{Ye_msp_2019}. In addition, the decay timescale of isolated young pulsars has negligible effects on the evolution of MSPS and, thus, is not essential for the discussion of this study.} On the other hand, during mass accretion in a binary, a pulsar's magnetic field decreases inversely proportional to the mass accreted as $(1+\Delta M/10^{-6}\,\msun)^{-1}$ according to the `magnetic-field burying' scenario \citep[e.g.,][and references therein]{Bhattacharya_vandenHeuvel_1991}, and the pulsar is spun up from angular momentum transfer \citep{hurley2002evolution,Kiel+2008}.

In addition to these standard assumptions, we have implemented in this study several updates into \texttt{CMC} for forming single MSPs as will be discussed in detail in Section~\ref{sec:pathways}.

\subsection{Modeling the Globular Cluster NGC~6752} \label{subsec:ngc6752}
For comparisons to pulsar and cluster observations, we focus in particular on one cluster, NGC~6752. NGC~6752 is a \textit{typical} core-collapsed cluster in the Milky Way \citep[][2010 edition]{Harris_1996}, thus our comparisons to this particular cluster are a reasonable proxy for core-collapsed clusters in general. It hosts the second largest number of observed MSPs in core-collapsed clusters\footnote{M~15 and NGC~6266 have the same number of observed pulsars, but they are about twice as massive as NGC~6752 and are more challenging to simulate. NGC~6624 hosts the most observed MSPs among core-collapsed clusters and is very close to the Galactic center. Its evolution is strongly affected by the Galactic tidal field, also making it more challenging to simulate than NGC~6752. The numbers of observed pulsars are taken from the GC pulsar catalog, and estimates of cluster properties are from https://people.smp.uq.edu.au/HolgerBaumgardt/globular/.}, with $9$ observed MSPs, where $8$ of them are single, and one is in a binary (GC Pulsar Catalog). To simulate this cluster, we adopt the initial conditions of the cluster model that has shown good agreement with the observations of NGC~6752 from \citet{Kremer+2020catalog}. Our simulations all have an initial number of stars $N=800000$, initial virial radius $R_v = 0.5$~pc, metallicity $Z=0.0002$, and Galactocentric distance $R_g=8$~kpc. The initial mass function follows a standard Kroupa broken power law \citep{Kroupa2001}, and the stars are sampled between $0.08$ and $150~\msun$. The initial binary fraction is $5\%$ across all primary masses, and the secondary masses are drawn from a uniform distribution in the range $0.1-1$ of the primary mass \citep{Duquennoy_Mayor_1991}. We use a King profile \citep{King1966} with a concentration parameter $W_0=5$ for the initial mass density distribution.

\begin{figure}
\begin{center}
\includegraphics[width=\columnwidth]{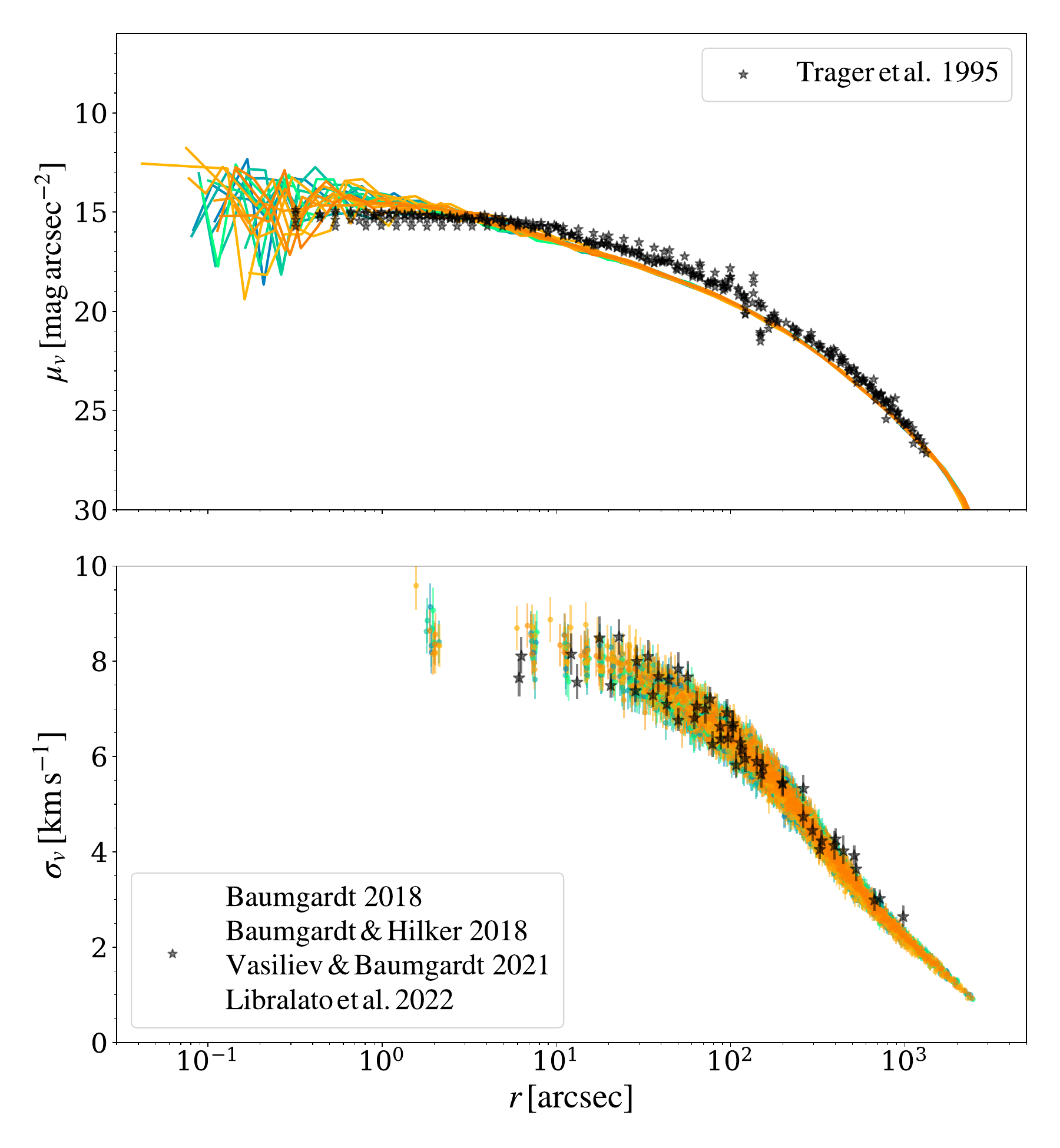}
\caption{Comparisons between model and observed surface brightness profiles and velocity dispersion profiles of the cluster NGC~6752. The observations are shown as black stars. The observed surface brightness profile is from \citet{Trager1995}, and the observed velocity dispersion profile is from https://people.smp.uq.edu.au/HolgerBaumgardt/globular/ and \citet[][line-of-sight radial velocities]{Baumgardt_2017}, \citet[][line-of-sight radial velocities]{Baumgardt_Hilker2018}, \citet[][proper motions]{Vasiliev+2021}, and \citet[][proper motions]{Libralato+2022}. The curves and circles with various colors show the profiles of different models at about 12~Gyr.}\label{fig:sbp_vdp_6752}
\end{center}
\end{figure}

%\startlongtable
%\newgeometry{left=0.5in}
\begin{deluxetable*}{c|cccccccccccccccc|cccc}
\tabletypesize{\scriptsize}
\tablewidth{-1pt}
\setlength{\tabcolsep}{1.5pt}
\tablecaption{Summary of Simulation Results} \label{tab:clu_prop}
\tablehead{
\colhead{Model} & \colhead{$M_{clu}$} & \colhead{$N_{\rm{BH}}$} & \colhead{$N_{\rm{NS}}$} & \colhead{$N^t_{\rm{WDc}}$} & \colhead{$N^t_{\rm{WDm}}$} & \colhead{$N^{\rm{NS}}_{\rm{TDE}}$} & \colhead{$N^s_{\rm{PSR}}$} & \colhead{$N^b_{\rm{PSR}}$} & \colhead{$N^s_{\rm{MSP}}$} & \colhead{$N^b_{\rm{MSP}}$} & \colhead{$N^s_{\rm{WDc}}$} & \colhead{$N^b_{\rm{WDc}}$} & \colhead{$N^s_{\rm{WDm}}$} & \colhead{$N^b_{\rm{WDm}}$} & \colhead{$N_{lmco}$} & \colhead{$r_{ib}$} & TDE & NSsp & WDTC & GCTC\\
\colhead{} & \colhead{$10^5\msun$} & \colhead{} & \colhead{} & \colhead{} & \colhead{} & \colhead{} & \colhead{} & \colhead{} & \colhead{} & \colhead{} & \colhead{} & \colhead{} & \colhead{} & \colhead{} & \colhead{} & \colhead{} & \colhead{} & \colhead{} & \colhead{} & \colhead{}
}
\startdata
1a & 1.97 & 5 & 968 & 88 & 2 & 0 & 7 & 10 & 5 & 9 & 40 & 2 & 0 & 0 & 1 & 4.7 & - & - & - & -\\
2a & 2.03 & 2 & 950 & 75 & 4 & 0 & 6 & 9 & 4 & 8 & 46 & 1 & 2 & 0 & 2 & 7.4 & $\checkmark$ & - & - & -\\
3a & 2.07 & 6 & 960 & 77 & 4 & 160 & 78 & 10 & 77 & 9 & 31 & 2 & 2 & 0 & 1 & 10.9 & $\checkmark$ & 0.2 & - & -\\
4a & 1.90 & 4 & 933 & 92 & 3 & 279 & 110 & 33 & 109 & 33 & 41 & 2 & 0 & 0 & 6 & 5.2 & $\checkmark$ & 0.2 & - & $\checkmark$\\
5a & 1.99 & 4 & 893 & 165 & 2 & 257 & 115 & 31 & 114 & 31 & 69 & 2 & 0 & 0 & 10 & 8.3 & $\checkmark$ & 0.2 & $\checkmark$ & $\checkmark$\\
6a & 2.01 & 3 & 889 & 138 & 3 & 166 & 79 & 11 & 76 & 11 & 70 & 3 & 1 & 0 & 2 & 12.9 & $\checkmark$ & 0.2 & $\checkmark$ & -\\
7a & 2.03 & 3 & 952 & 77 & 6 & 167 & 18 & 10 & 16 & 10 & 29 & 2 & 1 & 0 & 5 & 8.3 & $\checkmark$ & 0.8 & - & -\\
8a & 2.05 & 4 & 982 & 139 & 2 & 0 & 12 & 12 & 8 & 11 & 71 & 4 & 1 & 0 & 4 & 7.0 & $\checkmark$ & - & $\checkmark$ & -\\
$9a^*$ & 1.96 & 2 & 1238 & 70 & 3 & 0 & 24 & 9 & 19 & 9 & 41 & 0 & 2 & 0 & 2 & 9.7 & - & - & - & -\\
\hline
1b & 1.99 & 7 & 917 & 70 & 154 & 0 & 9 & 14 & 8 & 13 & 15 & 2 & 32 & 3 & 6 & 4.8 & - & - & - & -\\
2b & 2.02 & 9 & 970 & 62 & 151 & 0 & 8 & 11 & 5 & 9 & 17 & 1 & 37 & 1 & 3 & 6.6 & $\checkmark$ & - & - & -\\
3b & 1.99 & 10 & 879 & 79 & 124 & 152 & 62 & 21 & 60 & 21 & 19 & 3 & 29 & 1 & 6 & 6.1 & $\checkmark$ & 0.2 & - & -\\
4b & 1.99 & 4 & 878 & 85 & 182 & 287 & 107 & 25 & 106 & 24 & 15 & 1 & 36 & 2 & 6 & 7.7 & $\checkmark$ & 0.2 & - & $\checkmark$\\
5b & 1.95 & 9 & 840 & 121 & 172 & 270 & 72 & 31 & 72 & 31 & 17 & 3 & 36 & 1 & 11 & 5.6 & $\checkmark$ & 0.2 & $\checkmark$ & $\checkmark$\\
6b & 1.97 & 12 & 904 & 116 & 134 & 236 & 76 & 10 & 75 & 9 & 21 & 2 & 30 & 1 & 1 & 11.4 & $\checkmark$ & 0.2 & $\checkmark$ & -\\
7b & 2.06 & 9 & 971 & 54 & 137 & 167 & 9 & 11 & 7 & 10 & 7 & 2 & 37 & 2 & 4 & 6.1 & $\checkmark$ & 0.8 & - & -\\
8b & 2.04 & 7 & 916 & 103 & 145 & 0 & 9 & 12 & 8 & 11 & 28 & 3 & 43 & 2 & 3 & 6.3 & $\checkmark$ & - & $\checkmark$ & -\\
$9b^*$ & 2.01 & 15 & 955 & 106 & 93 & 192 & 86 & 19 & 84 & 19 & 20 & 3 & 16 & 1 & 5 & 7.4 & $\checkmark$ & 0.2 & $\checkmark$ & -

\enddata
\tablecomments{From left to right: Cluster mass, number of BHs, number of NSs, total number of WD--WD collisions where the collision remnant is an NS, total number of WD-WD mergers where the remnant is an NS, total number of NS--main-sequence star TDEs, number of single pulsars including both young radio pulsars and MSPs, number of binary pulsars including both young radio pulsars and MSPs, number of single MSPs, number of binary MSPs, number of single NSs formed in WD--WD collisions, number of NSs in binaries formed in WD--WD collisions, number of single NSs formed in WD--WD mergers, number of NSs in binaries formed in WD--WD mergers, and number of MSPs with very low-mass ($\lesssim0.01\,\msun$) CO/ONe companions. $r_{ib}$ is the ratio between single and binary MSPs, assuming that NSs from WD--WD collisions/mergers are born MSPs, and the MSPs with very low-mass companions are single MSPs. \newline ---All numbers are the averages over multiple model time steps at the present-day between 11 and 13.8~Gyr except $N^t_{\rm{WDc}}$, $N^t_{\rm{WDm}}$, and $N^{\rm{NS}}_{\rm{TDE}}$. These three values are counted throughout the simulations. \newline ---The last four columns indicate the specific dynamical processes that are included in the simulations. `TDE'--all compact object TDEs as described in \citet{Kremer+2019tde} and Section~\ref{subsec:nstdes}. `NSsp'-- mass transport efficiency from Eq.~\ref{eq:macc} in Section~\ref{subsec:nstdes} for the spin-up of NSs through TDEs. `WDTC'--WD--WD tidal captures. `GCTC'--binary formation through collisions with giant stars and tidal capture interactions (which does not include WD--WD tidal captures; \citealp[See][for more details]{Ye_47tuc_2021}). \newline ---The asterisks mark the models that use a different critical mass ratio prescription for unstable mass transfers in Section~\ref{subsec:bin_disrupt}.}
\end{deluxetable*}

In total, we have run $18$ NGC~6752-like simulations with the same initial conditions while allowing the new implementations specific for forming single MSPs (Section~\ref{sec:pathways}) to be varied. All simulations are run for $13.8$~Gyr. We compare all model surface brightness profiles and velocity dispersion profiles at about $12$~Gyr to the observations in Figure~\ref{fig:sbp_vdp_6752} (the observed cluster age is about 11-14~Gyr; \citealp{Buonanno+1986,Gratton+1997,Gratton+2003,Correnti+2016,Souza+2020,Bedin+2023}). We adopt $4.125$~kpc as the distance to the cluster \citep{Baumgardt_Vasiliev_2021}. Note that \citet{Vasiliev+2021} estimated that the semi-major axis of NGC 6752's orbit in the Milky Way is about 4.5~kpc with a small eccentricity of about 0.25, different from the Galactocentric distance we adopt here. This difference mostly affects the tidal boundary of the cluster and is not likely to affect the dynamical interactions in the cluster core which determine the number of single MSPs formed. As shown, all $18$ models closely match the observed profiles without fine-tuning of the cluster's initial conditions.

\section{Pathways for forming single millisecond pulsars}\label{sec:pathways}

In this section we discuss the various mechanisms for forming single MSPs in GCs and our prescriptions implemented for the first time in \texttt{CMC} for handling these pathways. Critically, since these pathways all involve rare events relevant specifically to NSs, they have a negligible effect on the global properties of their host cluster. As a result, all cluster simulations in our study are good fits for NGC~6752, independent of the differences pertaining to the MSP populations. Table~\ref{tab:clu_prop} shows the detailed properties and results of the models, to be discussed in detail in the following subsections. The models are separated into two sets, ‘a’ (using the default mass prescription) and ‘b’ (using the updated mass prescription), where set b forms more massive NSs as will be discussed in Section~\ref{subsec:wdtctde}. Models 1a and 1b have all mechanisms related to forming more single NSs and MSPs turned off and are the fiducial models.

\subsection{Disruption of Neutron Star Binaries}\label{subsec:bin_disrupt}
Perhaps the most straightforward way to form single MSPs is through binary-mediated dynamical encounters where the NS binaries that passed previously through a low-mass X-ray binary mass transfer phase are disrupted dynamically. In addition, MSP binaries may also be disrupted through the evaporation of the companion stars. Very low-mass ($\lesssim0.01\,\msun$) CO/ONe WD companions of some MSP binaries may be unstable and may evaporate, leaving behind single MSPs \citep{Bildsten_2002}. At the present day, $\sim 10$ single MSPs formed from these two channels are found per time step (models 1a and 1b in Table~\ref{tab:clu_prop} by adding $N_{lmco}$ to $N_{\rm{MSP}}^s$). The number alone can roughly match the number of observed single MSPs in NGC~6752, but the number of potentially observable MSPs is almost certainly much larger than the number we know now, given the sensitivity of radio telescopes and the MSP selection effects (also see Section~\ref{sec:discuss}). In addition, there are about similar numbers of binary MSPs in the models 1a and 1b (Table~\ref{tab:clu_prop} by subtracting $N_{lmco}$ from $N_{\rm{MSP}}^b$), and the extreme ratio of single-to-binary MSPs in NGC~6752 and some other core-collapsed GCs (Figure~\ref{fig:fsmap_compare} and Table~\ref{tab:obs_npsr}) is not explained. Here we further discuss the dynamical disruption of MSP binaries.

The rate of dynamical disruption of NS binaries is determined by the density of the host cluster \citep[often defined using the so-called $\Gamma$ parameter; e.g.,][]{Verbunt_Freire_2014} and the sizes of the NS binaries themselves which set their cross sections for encounters. A key parameter that determines the cross sections of these binaries is the stability criterion of binary mass transfer. The stability of mass transfer during Roche-lobe overflow in our simulations is determined by the critical mass ratio, $q_{crit} = M_{\rm{donor}}/M_{\rm{accretor}}$. Smaller $q_{crit}$ allows for unstable mass transfer from lower-mass donors and a wider range of donor-to-accretor mass ratios, which essentially corresponds to higher instability during mass transfer. In turn, this leads to higher rates of the common-envelope phase (note that we assume there is no mass accretion during the common-envelope process) and the formation of more compact NS--WD binaries. Indeed, in our previous study \citet{Ye_msp_2019}, many MSPs were spun up at least partially by a WD binary companion. These compact NS--WD binaries have very small encounter cross sections and are difficult to disrupt dynamically. On the other hand, higher values for $q_{crit}$ lead in general to more stable mass transfer which allows NSs to be spun up during accretion from a giant-branch companion, leading to a population of relatively wider MSP--WD binaries with larger cross sections for subsequent dynamical encounters that may break apart the binaries.

\begin{figure}
\begin{center}
\includegraphics[width=\columnwidth]{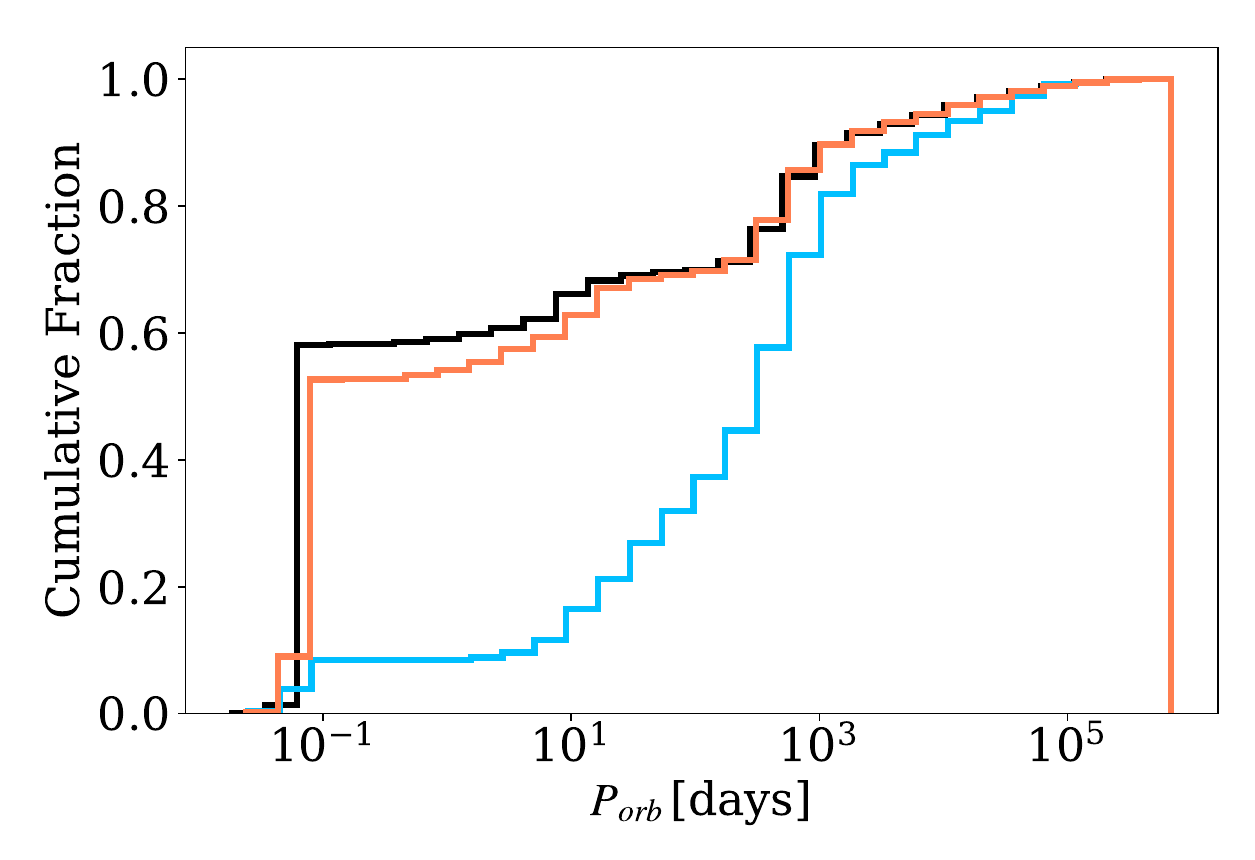}
\caption{Normalized cumulative distributions of orbital periods of MSP binaries at a Hubble time from three \texttt{COSMIC} population synthesis runs. All initial conditions of the three runs are the same except for the critical mass ratio, $q_{crit} = M_{\rm{donor}}/M_{\rm{accretor}}$, which controls the onset of a common envelope or unstable mass transfer during Roche lobe overflow \citep{cosmic}. The black histogram follows the prescriptions from \citet{hurley2000comprehensive} and \citet{Hjellming_Webbink_1987}, and the coral histogram follows \citet[][their Section~2.3]{Neijssel+2019}. The blue histogram uses the prescriptions in \citet[][their Section~5.1]{Belczynski+2008}, which in general has larger critical mass ratios for stars at the giant branch or the asymptotic giant branch than the other two prescriptions (see text for more detail).}\label{fig:qcrit_comp}
\end{center}
\end{figure}

To demonstrate the effects of $q_{crit}$ on the orbital separations of MSP binaries, we have run three NS isolated binary populations using \texttt{COSMIC} \citep{cosmic}. The initial mass function, metallicity, and the initial binary properties are sampled the same way as in the cluster simulations (Section~\ref{sec:method}). All three binary simulations have the same initial conditions, including the number of binaries, and we vary only the prescriptions for $q_{crit}$. We select the population of interest at the time of their formation. Figure~\ref{fig:qcrit_comp} compares the orbital period distributions of MSP binaries for three different $q_{crit}$ prescriptions. The black, coral, and blue histograms each contain $3936$, $261$, and $2992$ MSP binaries at a Hubble time, respectively. In general, the critical mass ratios from \citet{Belczynski+2008} (blue distribution) are larger for stars at the giant branch or the asymptotic giant branch (when the core mass of a giant is $\lesssim90\%$ of the giant's total mass), which lead to a much smaller fraction of tight MSP binaries. In the dense, dynamical environments of GCs, larger fractions of wide binaries may lead to more single MSPs since wide binaries have larger cross sections and more frequent dynamical encounters; thus, the original MSP binaries may be more easily disrupted and leave behind single MSPs.

To test the role of this mechanism in clusters, we run two cluster simulations using the critical mass ratio prescriptions from \citet{Belczynski+2008} (model 9a and 9b in Table~\ref{tab:clu_prop}). All other simulations use the prescriptions from \citet{hurley2002evolution} and \citet{Hjellming_Webbink_1987}, as was adopted in our previous paper \citet{Ye_msp_2019}. Model 9a has about four times more single MSPs than model 1a (not including single MSPs from WD--WD coalescences, although model 9b only has about the same single MSPs as model 6b), but is still not able to reproduce the observed ratio of single-to-binary MSPs. This suggests that the critical mass ratio for unstable mass transfer may play a role in forming single MSPs in GCs, even though the evolution of binaries in core-collapsed clusters is likely strongly influenced by dynamical encounters.

\subsection{NS Spin-up from TDEs of Main-sequence Stars} \label{subsec:nstdes}
\begin{figure*}
\begin{center}
\includegraphics[width=\textwidth]{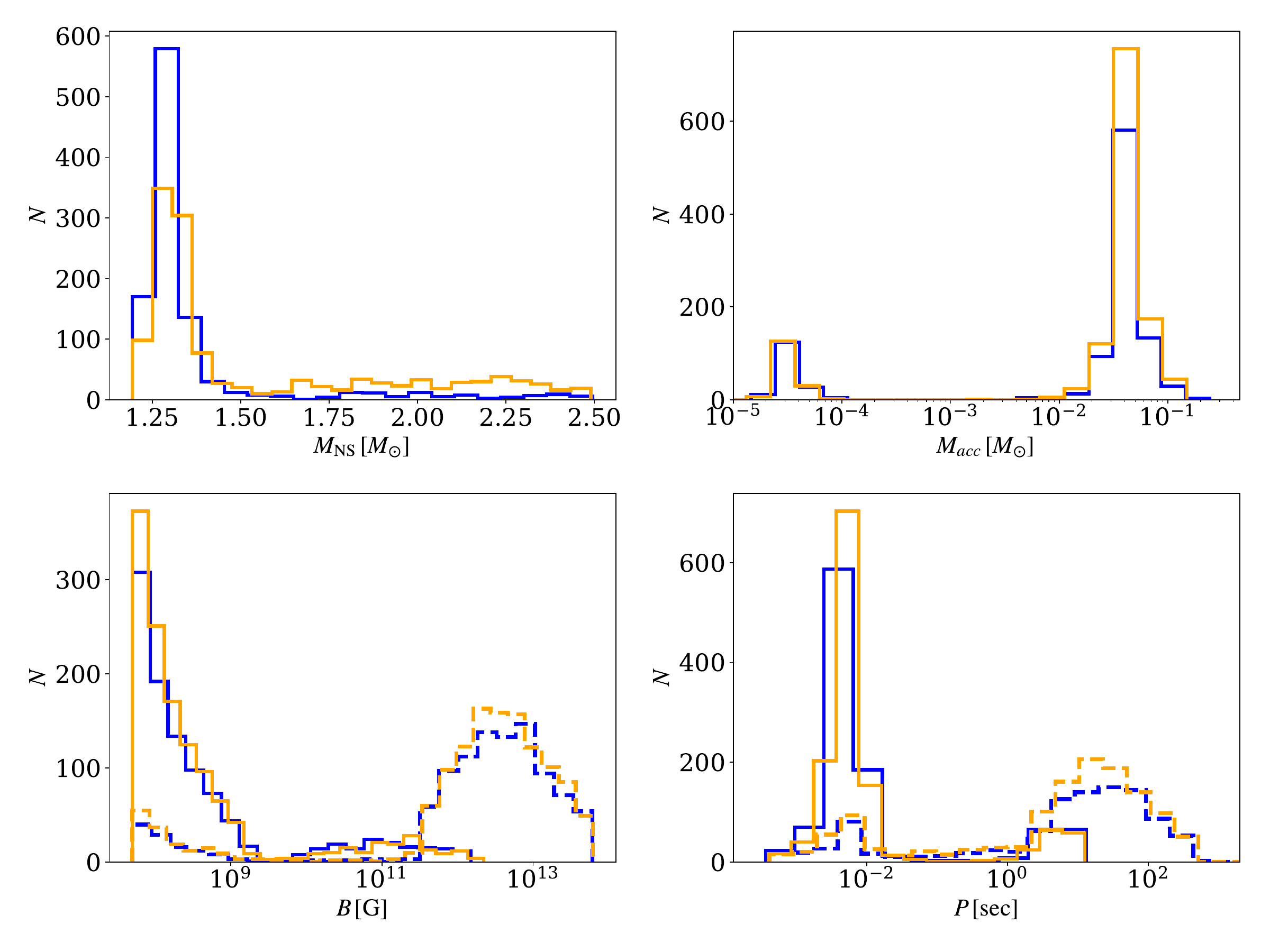}
\caption{Properties of NSs that went through TDEs with main-sequence stars. Blue and orange histograms are for the model sets a, and b, respectively. Model set a uses the default mass prescription and does not conserve mass during dynamical mass transfer in double WD binaries or during the collapses of heavy WDs, while model set b uses the updated prescription and conserves masses in both cases, thus producing more massive NSs. Upper left panel: the masses of NSs after the TDEs. Upper right panel: the amounts of mass accreted onto the NSs. The larger $M_{acc}$ peaks correspond to a higher accretion efficiency $s=0.2$, and the peaks at smaller $M_{acc}$ correspond to $s=0.8$ (see also Section~\ref{subsec:nstdes}). Lower left panel: magnetic fields of the NSs before (dashed histograms) and after (solid histograms) the TDEs. Lower right panel: NS spin periods before and after the TDEs. A high accretion efficiency during the TDEs leads to the spin-up of almost all NSs, while a low accretion efficiency can only spin up a few NSs.}\label{fig:tdeprop}
\end{center}
\end{figure*}

Second, NSs could be spun up through disruptions of main-sequence stars and accretion of the debris, producing single MSPs directly \citep{Kremer_nstde_2022}. This phenomenon would most frequently occur in core-collapsed GCs. \texttt{CMC} self-consistently tracks close encounters between NSs and main-sequence stars and computes the tidal disruption process.

The accretion of the debris following the tidal disruption of a main-sequence star may spin up an NS to millisecond periods \citep[e.g.,][]{Davies+1992,Lee+1996,Kremer_nstde_2022}. NS-main sequence TDEs can occur during close encounters with pericenter distance 
\begin{equation}
    r_p < \left(\frac{M_{\rm{NS}}}{M_*}\right)^{1/3}R_*\,,
\end{equation} 
where $M_{\rm{NS}}$ is the initial mass of the NS, and $M_*$ and $R_*$ are the initial mass and radius of the main-sequence star, respectively. For this study, we adopt the prescriptions in \citet{Kremer_nstde_2022} for spinning up NSs through TDEs.
    
The mass accretion rate is calculated as
\begin{equation}\label{eq:macc}
    \begin{aligned}
        \dot{M}_{acc}&\approx\left(\frac{M_d}{t_v}\right)\left(\frac{R_{acc}}{R_d}\right)^s\\
        &\times\left[1+3(1-C)\left(\frac{t}{t_v}\right)\right]^{-\frac{1+3(1+2s/3)(1-C)}{3(1-C)}}\,,
    \end{aligned}
\end{equation}
where $M_d$, $R_d$, and $t_v$ are the initial disk mass, disk radius, and viscous accretion time, respectively. We set $M_d = 0.9M_*$ following the hydrodynamic simulations of \citet{Kremer_nstde_2022} where it is shown that, in general, $\sim90\%$ of the initial stellar mass is bound to the NS following these TDEs. This is a reasonable assumption for encounters in GCs where the relative velocity of a pair of objects is in general much less than the stellar escape velocity. We assume that $R_d=2r_p$, $t_v=1~$day, and $R_{acc}$ equals the radius of the NS \citep{Kremer_nstde_2022}. The exponent $s \in [0,1]$ is a free parameter \citep[e.g.,][]{Blandford_Begelman_1999} defining the amount of material transported from the outer edge of the disk to the NS. This parameterization accounts for the fact that in these highly super-Eddington disks, an uncertain fraction of disk material is expected to be launched from the disk via winds \citep[e.g.,][]{Metzger2008, Kremer_nstde_2022}. We adopt $C=2s/(2s+1)$, equivalent to the assumption that the disk wind outflow produces no net torque on the disk \citep[e.g.,][]{Kumar2008}.
    
The accretion rate of the spin angular momentum is estimated as
\begin{equation}\label{eq:jacc}
    \dot{J}_{acc}\approx\dot{M}_{acc}\left(\sqrt{GM_{\rm{NS}}R_{acc}}+\frac{1}{2}M_{acc}\sqrt{\frac{GR_{acc}}{M_{\rm{NS}}}}\right)\,,
\end{equation}
where $M_{acc}$ is the total mass accreted onto the NS.
    
We analytically integrate Eq.~\ref{eq:macc}-\ref{eq:jacc} (Eq.~3-4 in \citealp{Kremer_nstde_2022}) for $10^6$~seconds to compute the total mass and angular momentum accreted onto the NS. During the TDE, the NS is spun up accordingly through angular momentum transfer \citep[][Eq. 54]{hurley2002evolution}, and its magnetic field is assumed to decay following the same `magnetic-field burying' scenario as in \citet[][their Eq. 3]{Ye_msp_2019} for mass accretion in a binary.

Figure~\ref{fig:tdeprop} shows the properties of all NSs in the simulations that tidally disrupted main-sequence stars \footnote{We include here also NSs formed in a WD--WD collision right before the TDE during one close encounter ($\lesssim 5\%$ of all TDEs).}.  The amounts of mass accreted onto the NSs depend on the mass inflow rate and can differ by $\sim3$~orders of magnitude between high ($s=0.2$) and low ($s=0.8$) efficiency (upper right panel and also Figure~3 in \citealt{Kremer_nstde_2022}). In the optimistic case where the inflow rate is high, almost all NSs that went through main-sequence star TDEs were spun up to MSPs (lower panels). This can boost the number of single MSPs in a typical core-collapse cluster by a factor of $\gtrsim 10$ (Table~\ref{tab:clu_prop}). On the other hand, low inflow rates increase the numbers by a factor of $\lesssim 3$. In addition, $11$ NSs in all relevant simulations exceeded the maximum NS mass after accreting from the debris of the disrupted main-sequence stars and collapsed to become BHs.

We want to point out that $\sim 10\%$ of the NS-main-sequence star TDEs include more than two objects in the close encounters. The outcome of these few-body interactions is uncertain and requires careful studies with hydrodynamics simulations. Here we allow these NSs to be spun up to MSPs to maximize the single MSPs formed in a model.

\begin{figure}
\begin{center}
\includegraphics[width=\columnwidth]{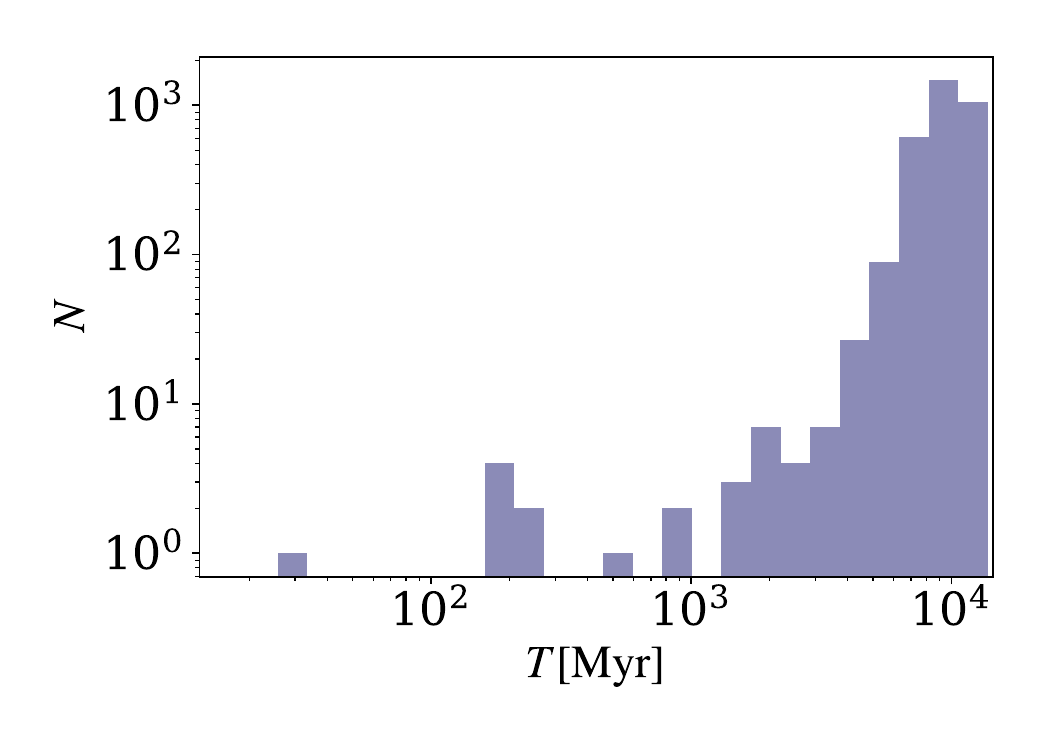}
\caption{The time of collisions (and TDEs if applicable) between NSs and main-sequence stars in all simulations. The number of close encounters peaks at $\sim 9$~Gyr.}\label{fig:t_nsms_coll}
\end{center}
\end{figure}

The average rate of NS--main-sequence star TDEs is about $60\,\rm{Gpc^{-3}\,yr^{-1}}$ at $\gtrsim 9$~Gyr (up to redshift $z\sim 0.5$) for typical core-collapsed clusters assuming a GC number density of $2.31~\rm{Mpc^{-3}}$. Figure~\ref{fig:t_nsms_coll} illustrates the time of collisions or TDEs between an NS and a main-sequence star in all simulations. If a fraction of energy is converted to luminosity during accretion, we can roughly estimate the luminosity to be $L_{\rm{TDE}} \sim \eta \dot{M_{acc}}c^2 $ which is $\sim 2\times10^{43}\,\rm{erg\,s^{-1}}$, assuming $\eta=0.01$ and $\dot M_{acc}\sim0.04\,\msun\,\rm{yr^{-1}}$ \citep{Kremer_nstde_2022}. 

In reality, the outcomes of these TDEs are uncertain and additional effects may limit the total mass growth by the NS. For instance, accretion feedback \citep[e.g.,][]{Armitage_Livio_2000,Papish+2015} or feedback due to nuclear burning at the NS surface \citep[e.g.,][]{Hansen_vanHorn_1975,Taam1985,Bilsten1998} may unbind the material initially bound to the NS and inhibit growth, provided the feedback energy can couple mechanically with the bound material. Our intent here is to explore whether these TDEs may account for the populations of observed single MSPs in the most optimal scenario where these feedback effects play a negligible role.

Finally, we also include WD--main-sequence star TDEs in \texttt{CMC}. Following the treatments in \citet{Kremer+2019tde}, we assume an immediate collision between a WD and a main-sequence star if their pericenter distance during the first passage is smaller than the tidal disruption radius of the main-sequence star, $(M_{\rm{WD}}/M_*)^{1/3}R_*$. The mass of the collision product is the total mass of the WD and the main-sequence star, so the product may evolve to be a more massive WD \citep{hurley2002evolution}, increasing the WD's interaction rates and the probability of subsequent double WD collisions and NS formation.

\subsection{White Dwarf Binary Coalescence and Collapse} \label{subsec:wdtctde}
A third way to form single MSPs is through the merger-induced collapse of WDs. To fully understand this channel, we have implemented tidal capture interactions between two WDs into \texttt{CMC}. During single-single or few-body interactions, if the pericenter distance of the approaching WDs is smaller than twice the sum of their radii (see Appendix~\ref{sec:append} for more details), we assume an immediate collision of the two WDs. This is a reasonable assumption since tidally captured double WD binaries are very compact and will merge in a short timescale (e.g., two solar-mass WDs captured at the maximum pericenter distance and eccentricity $e=0.99$ will merge in about ten thousand years, less than the typical timescale between dynamical encounters in a typical GC)\footnote{The crossing time during strong small $N$-body interactions is generally shorter than this inspiral timescale, but we allow the tidally captured WDs to merge immediately to maximize the double WD collision rate.}.

In this study, we focus on WD coalescences that can lead to the formation of NSs. These include both direct coalescences when two WDs physically collide (i.e., the pericenter distance of the encounter is smaller than the sum of the radii of the two WDs) and coalescences that follow more distant encounters leading to tidal capture and gravitational-wave inspiral. In addition to these dynamically-mediated scenarios, WD coalescences can also occur following unstable mass transfer in a gravitational-wave inspiralling WD--WD binary \citep[e.g.,][]{Webbink_1984,Nelemans2001,Marsh2004,Kremer+2021WD}. We assume unstable mass transfer ensues for WD--WD binaries with mass ratio $q>0.628$ \citep[e.g.,][]{cosmic}. Such binary mass transfer-mediated mergers are qualitatively similar to the physical collision and tidal capture encounters facilitated by close flybys, since in both cases the coalescence occurs on the dynamical timescale of the WDs involved. Thus, we assume that all of these scenarios result in the same outcome.\footnote{Perfectly head-on collisions may result in a unique outcome \citep[e.g.,][]{KatzDong2012}; however, since such events are relatively rare, we do not distinguish them here from the more general grazing collisions.}

\begin{figure}
\begin{center}
\includegraphics[width=\columnwidth]{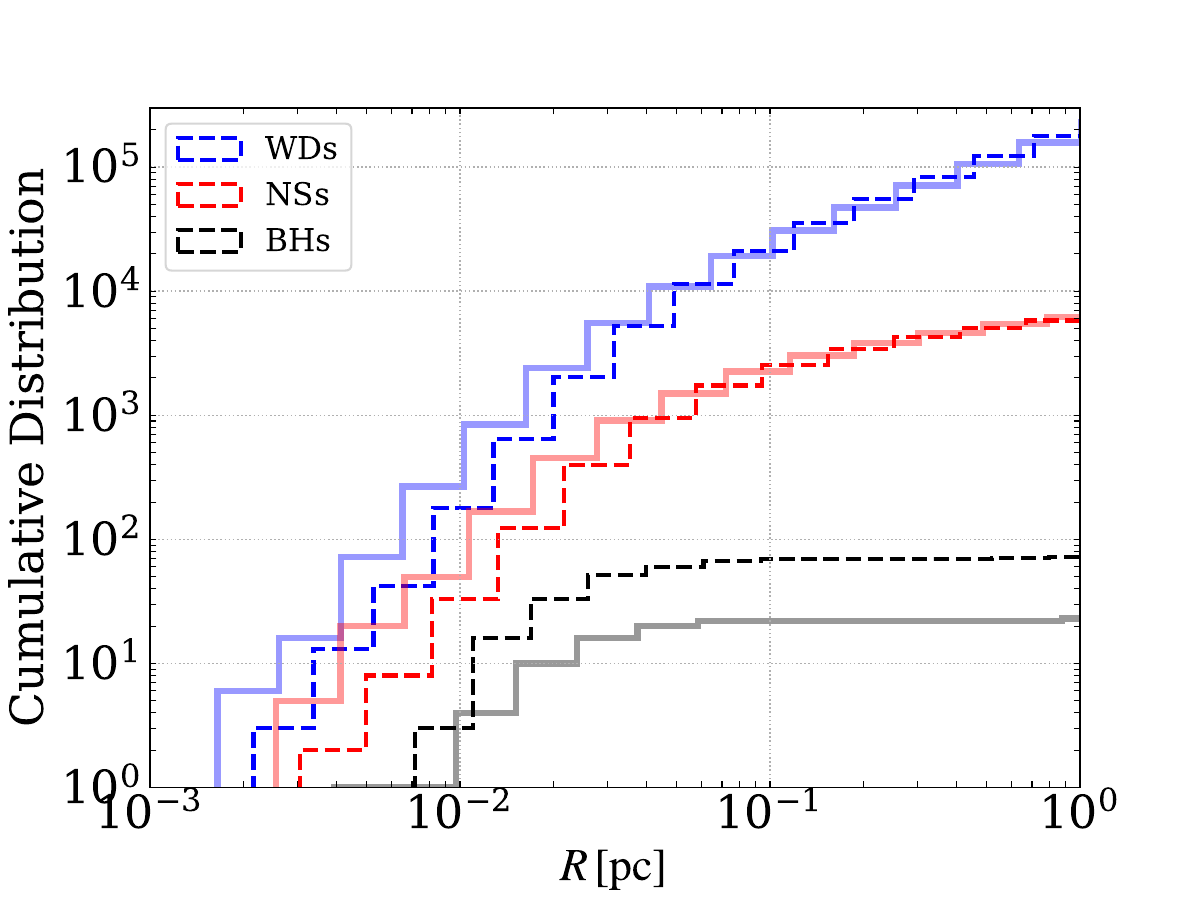}
\caption{Cumulative radial distributions of compact objects at about $12$~Gyr from all models. Histograms with lighter colors show the compact objects from model set a, which uses the default mass prescription and does not conserve masses during the mergers of heavy WDs and the collapses of the remnants, while the darker and dashed histograms show model set b, which uses the updated mass prescription and conserves masses during mergers and collapses, thus producing more massive NSs and more low-mass BHs.}\label{fig:rdistri}
\end{center}
\end{figure}

\begin{figure*}
\begin{center}
\includegraphics[width=\textwidth]{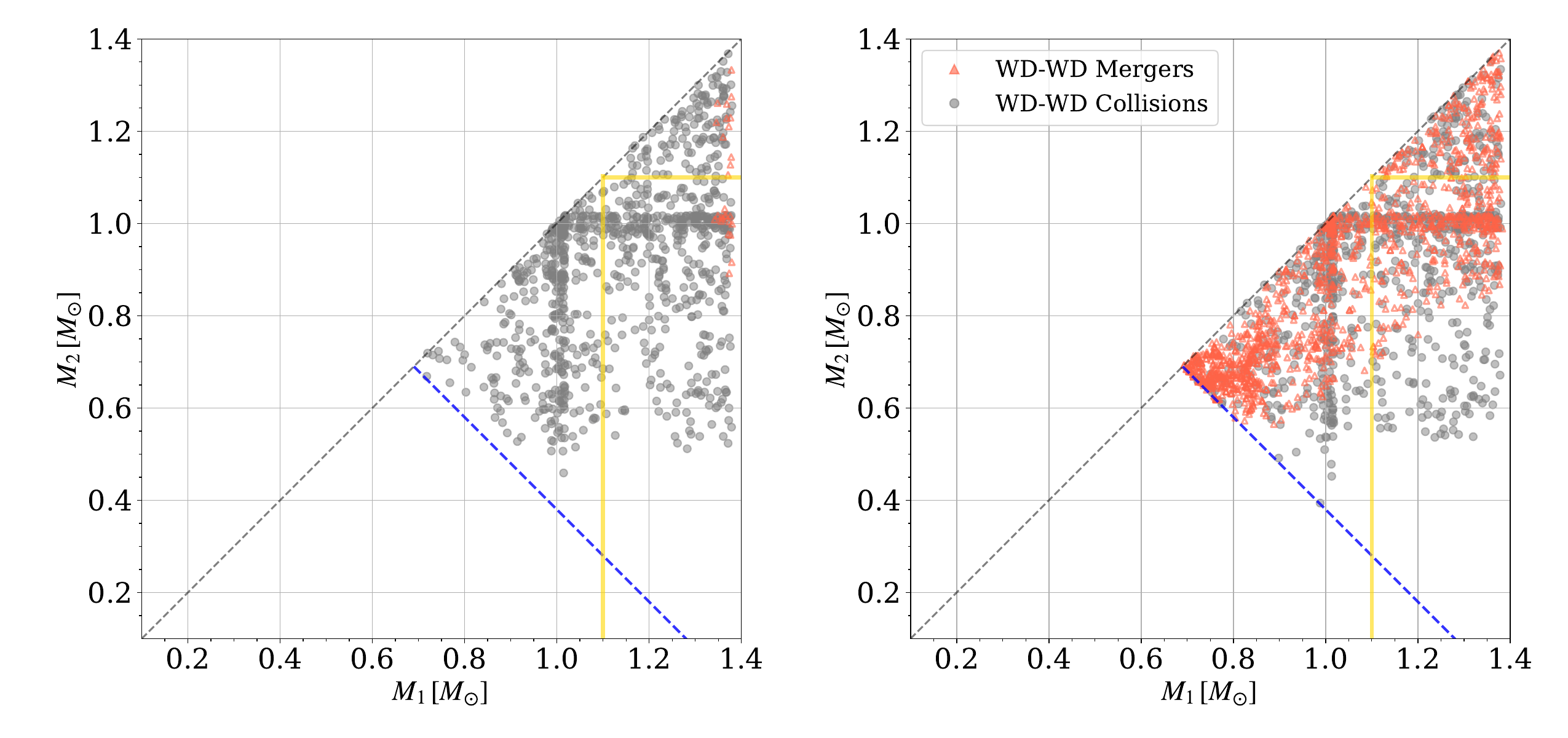}
\caption{Masses of component WDs in all super-Chandrasekhar WD--WD collisions or mergers in the simulations. Collisions refer to the dynamically-mediated coalescences following tidal capture and gravitational-wave inspiral or direct coalescences when the pair of WDs physically collides. Mergers occur following unstable mass transfer in double WD binaries. The left panel shows the masses from models 1-9a (which use the default mass prescription for dynamical mass transfer and merger-induced collapse of heavy WDs), and the right panel shows models 1-9b (which use updated mass prescription that conserves mass during the dynamical mass transfer and collapse of heavy WDs). The yellow lines mark $1.1\,\msun$, which is about the maximum mass of CO WDs at metallicity $Z=0.0002$. The blue dashed lines show where the total mass of the two interacting WDs equals $1.38\,\msun$.}\label{fig:wdmass}
\end{center}
\end{figure*}

Traditionally, coalescences of massive (i.e., super-Chandrasekhar) WD pairs have been connected to Type Ia supernovae \citep[e.g.,][]{Webbink_1984}. However, other studies have noted that in some cases, mechanisms such as runaway electron capture \citep[e.g.,][]{Nomoto_Iben_1985} or off-center carbon ignition leading to the formation of a low-mass iron core \citep[e.g.,][]{Schwab+2016} may allow a thermonuclear explosion to be avoided. In the latter cases, collapse to an NS is the likely result. Here we assume all super-Chandrasekhar coalescences where the components are CO WDs and/or ONe WDs lead to collapse \citep[e.g., also][]{Kremer+2023}, thus maximizing the formation rate of single NSs through this channel. The outcomes of various combinations of super-Chandrasekhar mass WD coalescences adopted in our simulations are shown in Table~\ref{tab:colloutcome}. We also list the assumed outcomes for cases where a WD accretes to the Chandrasekhar limit via stable Roche lobe overflow in a binary (assumed for mass ratios $q<0.628$), which is a distinct process from the dynamical-timescale coalescences.

%\startlongtable
\begin{deluxetable}{c c c c}
\tabletypesize{\scriptsize}
%\tablewidth{4pt}
%\setlength{\tabcolsep}{3pt}
\tablecaption{Outcomes of Super-Chandrasekhar WD--WD Interactions} \label{tab:colloutcome}
\tablehead{\colhead{WD Types} & \colhead{Collision} & \colhead{Dynamical MT} & \colhead{Stable MT}
}
\startdata
He+CO WD & He Star & He Star & SN Ia\\
He+ONe WD & He Star & He Star & NS\\
CO+CO WD & NS & NS & SN Ia\\
CO/ONe+ONe WD & NS & NS & NS\\
\enddata
\tablecomments{From left to right: The types of the two interacting WDs, the outcomes of collisions, the outcomes of dynamical mass transfers (MT) in binaries, and the outcomes of stable mass transfers. `He Star' means either naked helium star at the hertzsprung gap or naked helium star at the giant branch \citep{hurley2000comprehensive}. `SN Ia' is supernova type Ia. Note that because of the high mass ratios in CO+CO WD binaries, they likely do not go through stable mass transfer, but the outcome is included here for completeness.}
\end{deluxetable}

Following the dynamical coalescence of a massive WD pair \citep[e.g.,][]{Dan+2014}, subsequent phases of viscous and thermal evolution \citep[e.g.,][]{Shen+2012,Schwab2012} that can last as long as $10^5\,$yr and potentially include phases of dusty wind mass loss \citep[e.g.,][]{Schwab+2016} determine whether an NS forms and, if so, its final properties (e.g., mass, spin period, and magnetic field; \citealp{Schwab_2021}). Simulating in detail these various phases is beyond the scope of \texttt{CMC}, so in order to bracket some of the uncertainties inherent to these phases, we adopt two limiting assumptions in our simulations. As an upper limit (model set b in Table~\ref{tab:clu_prop}), we assume that the total mass loss during these phases is negligible so that the total mass of the WDs is conserved up through collapse. Accounting for the fractional loss of mass through gravitational-wave emission, we then assume the final masses of the NSs born following the collapses are $90\%$ of the initial WD pair. In practice, this choice serves to maximize the rate of single NS formation.

As an alternative lower-limit scenario, the other half of the simulations (model set a) instead follow the default prescriptions from \citet{cosmic}. In these simulations, the more massive WD accretes a total mass $\Delta M = \dot M_{\rm{edd}}\tau_{\rm{\dot M}}$ from the donor WD. Here $\dot M_{\rm{edd}}$ is the Eddington accretion limit and $\tau_{\rm{\dot M}} = \sqrt{\tau_{\rm{KH}}\tau_{\rm{dyn}}}$ is the characteristic mass-transfer timescale where $\tau_{\rm{KH}}$ and $\tau_{\rm{dyn}}$ are the Kelvin-Helmholtz and dynamical timescale of the donor WD, respectively. For WDs that are $\sim 1\,\msun$, $\dot M_{\rm{edd}}\approx10^{-5}\,\msun\,\rm{yr^{-1}}$, $\tau_{\rm{\dot M}}\approx 2000~$yr, and thus a very small amount of mass $\Delta M \approx 10^{-3}\,\msun$ is accreted \citep{cosmic,Kremer+2021WD}. The masses of all NSs formed in merger-induced collapses in these simulations (set a) are fixed to be $0.9\times1.38\,\msun$, the same as for NSs formed in accretion-induced collapses.

The number of WD--WD coalescences that collapse to NSs in each simulation is shown in Table~\ref{tab:clu_prop}.\footnote{Here we exclude the NSs which form from the collisions of more than two WDs.} Model set b differ from set a in: 1) the mass of the NSs formed in merger-induced collapse, with model set b having more massive NSs; and 2) the amount of mass accreted onto the primary WD during dynamical mass transfer, with models set b conserving the mass and having a larger number of NSs formed from dynamical mass transfer between two WDs. Note that even without additional dynamical interactions such as WD--WD tidal captures and NS spun-up through TDEs (Section~\ref{subsec:nstdes}), the number of single MSPs is boosted simply by assuming that dynamical mass transfer in double WD binaries and the subsequent collapses of the super-Chandrasekhar-mass merger products conserve mass (e.g., model 1b versus 1a, where there are more mergers in 1b). 

Overall, model set b (with the updated mass prescription) has more combined WD-WD coalescences that produce NSs than model set a (with the default mass prescription). This is largely because the conservation of mass during dynamical mass transfer in double WD binaries leads to many more NS remnants from mergers, despite model set b having fewer NS remnants from collisions. The decreased collisions are probably caused by mergers reducing part of the WDs available for collisions and the larger number of low-mass BHs leading to larger WD radial distributions through the energy release during BH dynamical encounters \citep[e.g.,][]{Kremer+2020bhburning}. Figure~\ref{fig:rdistri} illustrates the radial distributions of the compact objects within $1$~pc from the cluster centers. The WDs in the model set a (using the default mass prescription) are slightly more concentrated than those in model set b (using the updated mass prescription).

\begin{figure}
\begin{center}
\includegraphics[width=\columnwidth]{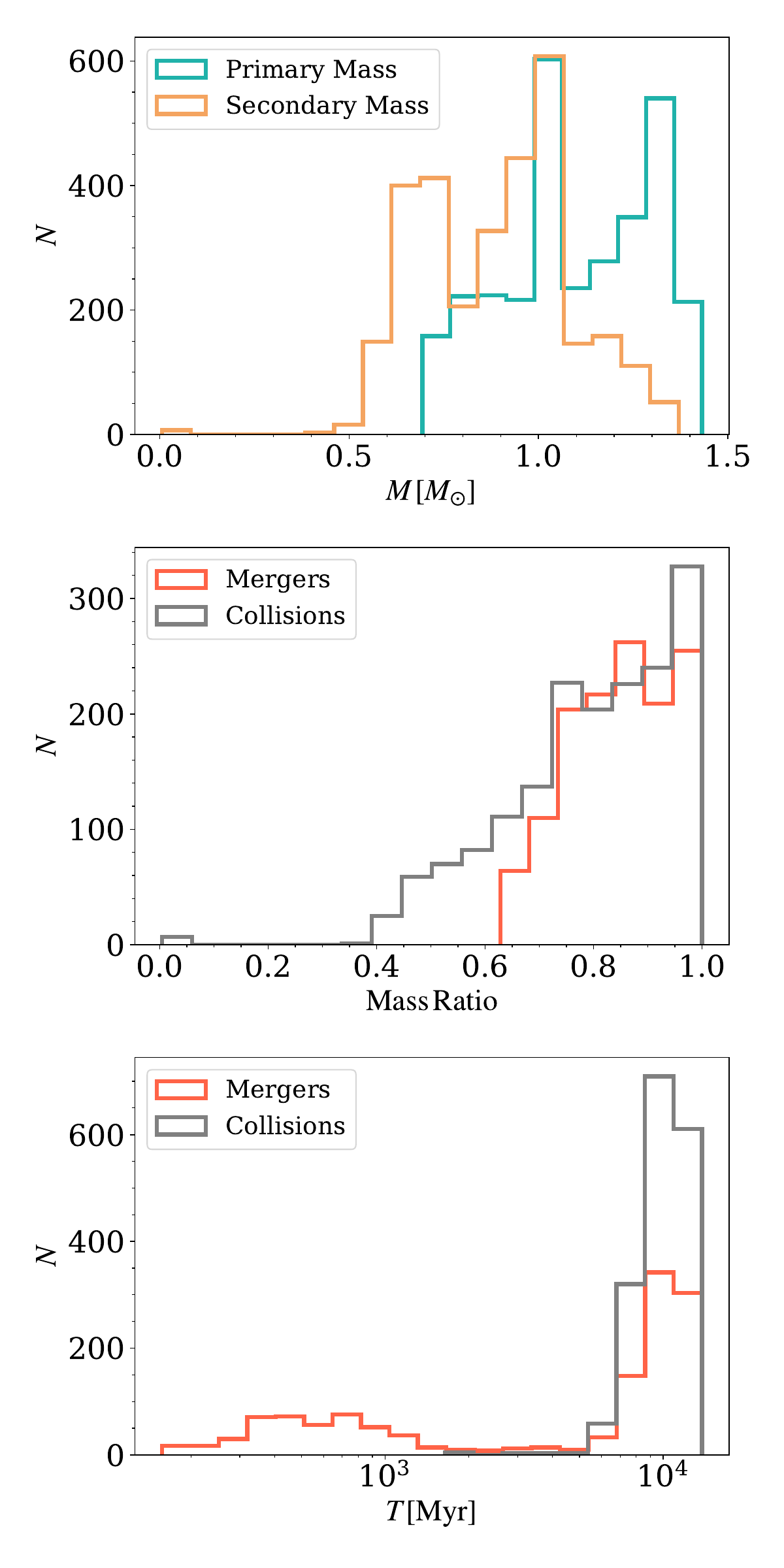}
\caption{Properties of all super-Chandrasekhar WD--WD mergers/collisions in the simulations. Upper panel: mass distributions of the primary and secondary WD masses. Middle panel: mass ratios of WDs in mergers and collisions. Bottom panel: distributions of the time at which mergers/collisions occur. The peak at $\lesssim 1$~Gyr is from the mergers of primordial double WD binaries, and the later peaks at $\sim 9$~Gyr are from the mergers of dynamically assembled double WD binaries or direct collisions during close encounters.}\label{fig:wdmass_distr}
\end{center}
\end{figure}

The masses of the component WDs involved in the coalescences are plotted in Figure~\ref{fig:wdmass}. Overall, CO WD+CO WD interactions account for $\sim 70\%$ of mergers and $\sim60\%$ of the collisions. The rest of the collisions or mergers include at least one ONe WD. The over-density of WDs at around $1\,\msun$ comes from the zero-age main-sequence (ZAMS) mass to WD mass relation we adopt for low metallicities  (\citealp{hurley2000comprehensive} and \citealp[Figure~1 in][]{Kremer+2021WD}). These WDs are the remnants of ZAMS stars of about $2.5-3.5\,\msun$.

We also show the distributions of component mass, mass ratio, and coalescence time of the WDs in Figure~\ref{fig:wdmass_distr}. In addition to the peaks at about $1\,\msun$ as mentioned above, the peak at roughly $0.6\,\msun$ arises from the initial mass function of ZAMS stars of about $1\,\msun$ \citep[also see][]{Kremer+2021WD}. Meanwhile, mass segregation of massive WDs and previous collisions/mergers contribute to the peak at around $1.3\,\msun$. 

Most mergers and collisions have mass ratios larger than $0.6$. The peak at around $1$ for collisions can be naturally explained by mass segregation and the more frequent close encounters of the massive WDs in the cluster cores (see also Figure~\ref{fig:rdistri}). Furthermore, the critical mass ratio for unstable mass transfer in double WD binaries ($0.628$ in all simulations) limits the mergers to a narrower mass ratio range than the collisions. 

Lastly, almost all WD--WD collisions and about $60\%$ of WD--WD mergers occur at late times ($\gtrsim 6$~Gyr) of the host clusters' evolution. At early times, BHs dominated the cluster cores because of mass segregation, while many WDs have yet to form. Frequent dynamical interactions eject many BHs from the clusters in a few billion years \citep[e.g.,][]{Kremer+2020bhburning} and the most abundant WDs come to dominate the cluster cores and engage in dynamical encounters at late times (see also Figure~\ref{fig:rdistri}). Almost all WD--WD binaries that merged at late times are dynamically assembled. In contrast, more than $90\%$ of the mergers at early times are from primordial binaries.

By assuming that all these WD--WD collisions and mergers produce MSPs directly (for alternatives, see Section~\ref{sec:discuss}), these two channels can also significantly increase the number of single MSPs in GCs (Table~\ref{tab:clu_prop}). The number of single MSPs formed in these channels is about half of those formed in TDEs (e.g., model 3a) or roughly comparable if WD--WD tidal capture is included or if masses are conserved during dynamical mass transfer in double WD binaries and merger-induced collapse of heavy WDs (e.g., model 6a and 3b). Thus the WD channels alone can also boost the number of single MSPs by a factor of $\sim 5-10$.

\section{Neutron Star Masses and Offsets}\label{sec:mass_offset}
\begin{figure}
\begin{center}
\includegraphics[width=\columnwidth]{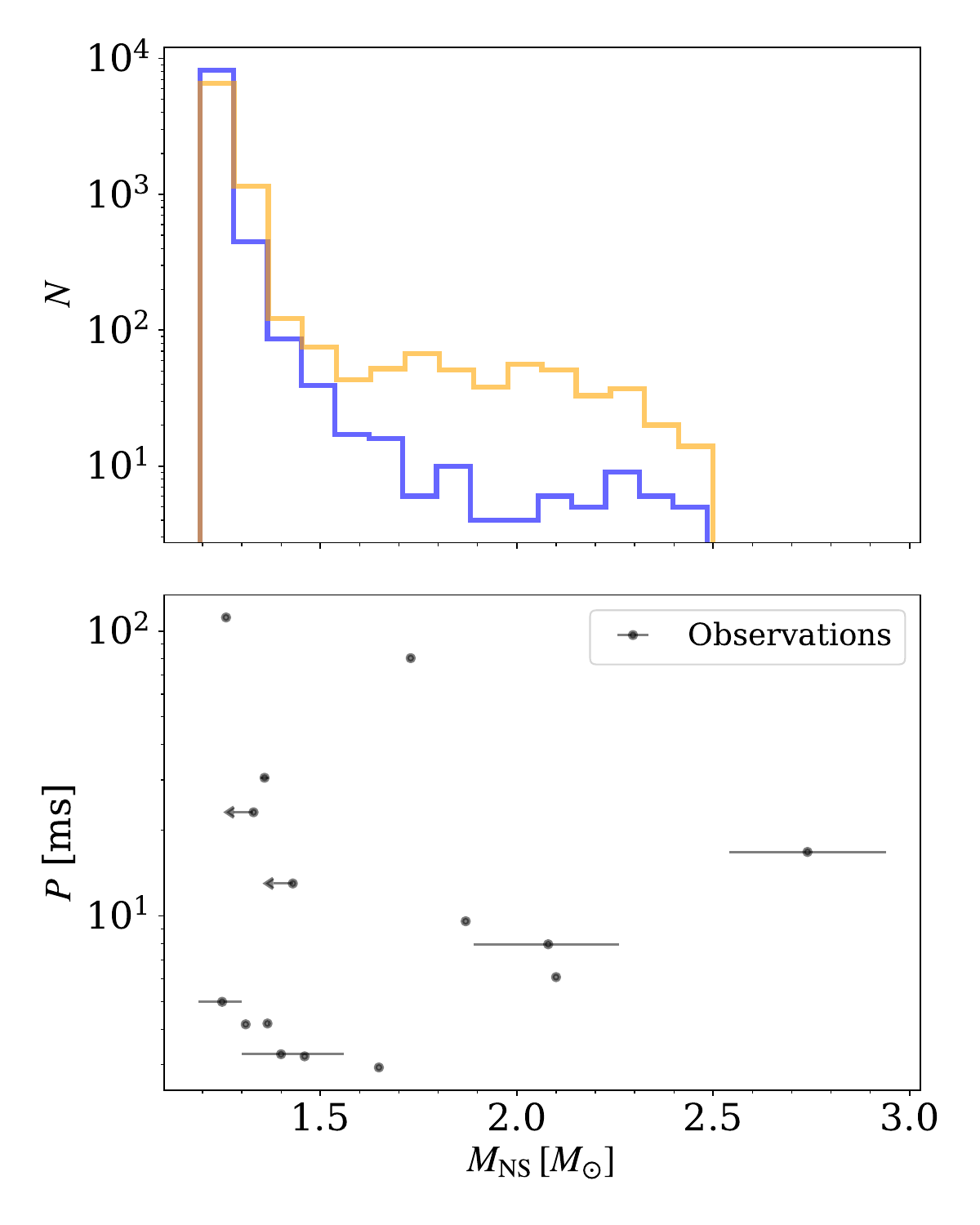}
\caption{Distributions of NS masses at about $12$~Gyr for model sets a (default mass prescription; blue histograms) and b (updated mass prescription; orange histograms) as in Table~\ref{tab:clu_prop} in the upper panel. The black dots in the lower panel show the known masses of pulsars from the GC Pulsar Catalog versus their spin periods \citep[e.g.,][]{Andersen_Ransom_2018,Bassa+2006,Begin_2006,Freire+2008ngc64s,Freire+2008M5,Freire+2017,Jacoby+2006,Lynch+2012,Ransom+2005,Ridolfi+2019,Ridolfi+2021,Thorsett_Chakrabarty_1999}. The observed masses without error bars are the median pulsar masses assuming a flat $cos\,i$ distribution where $i$ is the inclination angle of a pulsar binary. The mass and error bar of the most massive observed pulsar is the median assuming a flat $cos\,i$ distribution and its $1\sigma$ uncertainty, respectively.}\label{fig:nsmass}
\end{center}
\end{figure}

The masses of the NSs in all models at about $12$~Gyr are shown in Figure~\ref{fig:nsmass}. The mass distributions peak at $\sim1.2\,\msun$, which is the mass of NSs formed in accretion-induced collapses of WDs (see also Section~\ref{subsec:code}). Most of the massive NSs ($\sim 70\%$ when $\gtrsim1.5\,\msun$) in model set a (default mass prescription) come from collisions after the initial NSs were formed, where the initial NSs gain mass by merging with WDs or the dense cores of giant stars \citep{hurley2002evolution}. The rest were born massive in CCSNe or became massive through accretion. As is expected, model set b (updated mass prescription) has many more massive NSs than model set a from merger-induced collapses and dynamical mass transfer. The black stars in the figure show the observationally constrained cluster pulsar masses. The NS masses from both model sets are consistent with the observations. Future observations on pulsar masses would probably put better constraints on the formation of massive NSs. 

In addition, model set b also produces more low-mass BHs. The masses of these low-mass BHs are within the `lower-mass gap' between $\sim2.5$ and $\sim 5\,\msun$ inferred from observations of low-mass X-ray binaries \citep{Bailyn+1998,Ozel+2010,Farr+2011}. If merger-induced collapses of WDs could lead to the formation of low-mass BHs (or very massive NSs), they may potentially contribute to the observed `mass-gap' objects \citep{Thompson+2019,gw190814,Lam+2022}.

\begin{figure}
\begin{center}
\includegraphics[width=\columnwidth]{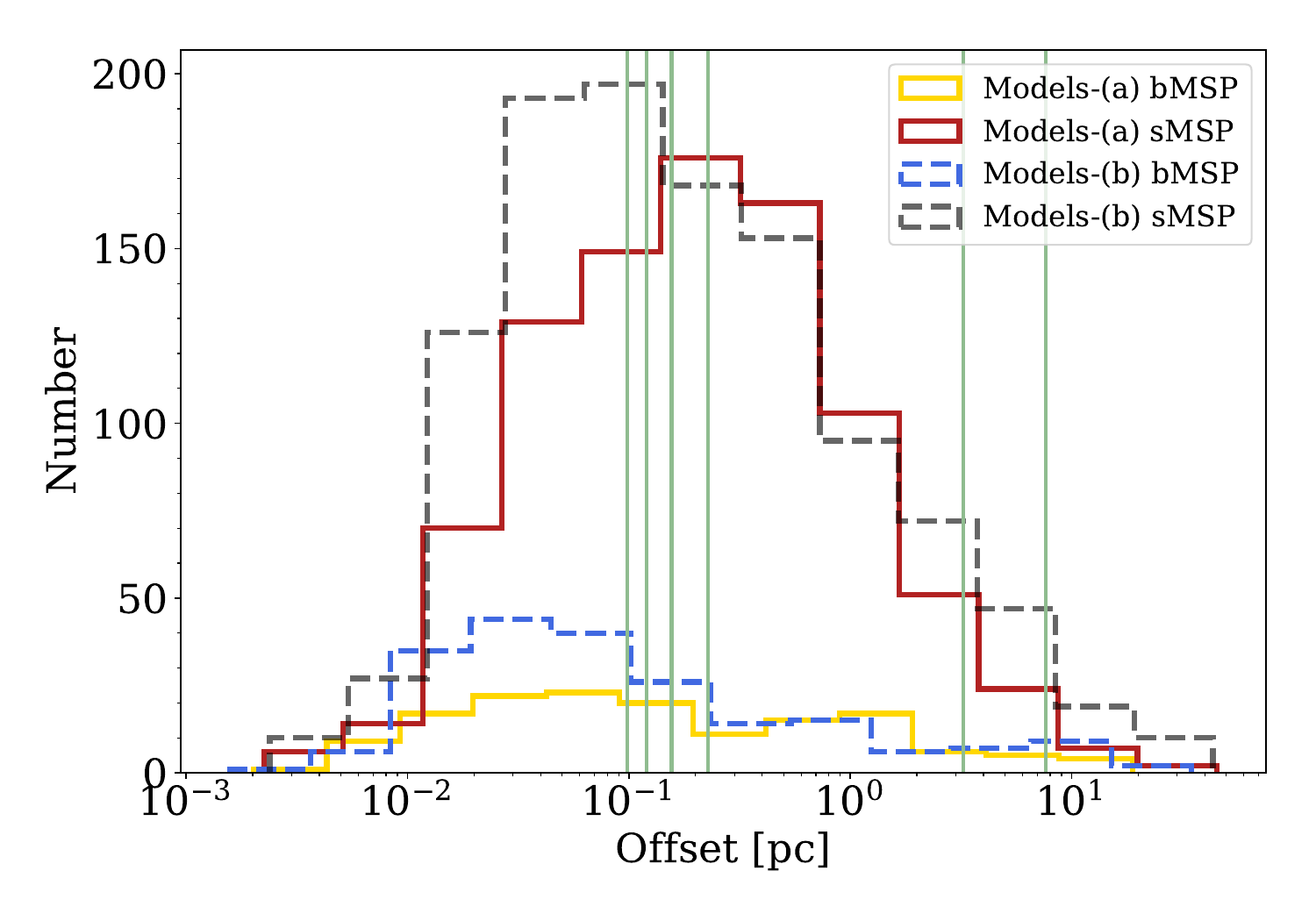}
\caption{Projected offset distributions of MSPs for all models in Table~\ref{tab:clu_prop} at about $12$~Gyr. The green vertical lines show the offsets of the observed pulsars in NGC~6752, assuming the cluster is at 4.125~kpc from Earth \citep{Baumgardt_Vasiliev_2021}. `bMSP' is MSPs in binaries, and `sMSP' is single MSPs. Models set a (solid histograms) and b (dashed histograms) are shown separately. The differences in the offset distributions of the two sets of models are small.}\label{fig:offset_6752}
\end{center}
\end{figure}

Mass differences may affect how concentrated the NSs are in the clusters because of mass segregation. Figure~\ref{fig:offset_6752} compares the offsets of the observed MSPs in NGC~6752 to those from model sets a (default mass prescription) and b (updated mass prescription). All the observed pulsars are single except for the outermost one. As expected, there are more massive MSPs in model set b than in model set a, so the MSP offsets peak at slightly smaller distances towards the cluster centers in model set b. Despite this minor difference, the peaks of the single MSPs for both model sets overlap with where most of the observed MSPs are. The distributions of the binary MSPs from the models also agree with the observed binary at the cluster outskirt, which may be ejected to the cluster halo through binary-single strong encounters \citep[e.g., see also][]{Leigh+2023}.

\begin{figure*}
\begin{center}
\includegraphics[width=\textwidth]{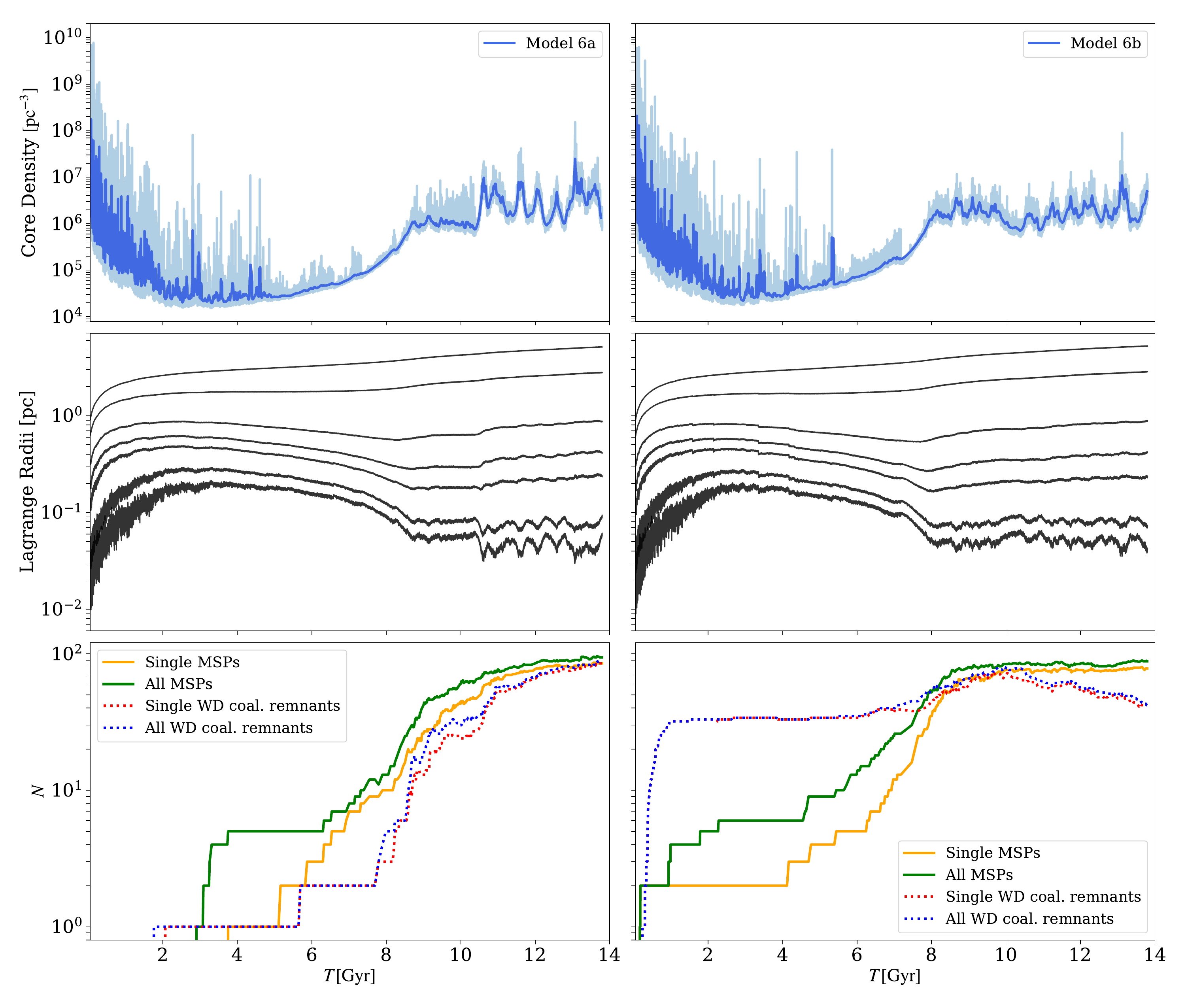}
\caption{The evolution as a function of time (in Gyr) of the core number density (upper panels), Lagrangian radii (middle panels), and the number of MSPs and the number of NS remnants from heavy WD coalescences (bottom panels) for two models (model 6a on the left and model 6b on the right). The darker blue curves in the upper panels show the smoothed core number densities of the original densities in lighter blue. The Lagrangian radii from bottom to top correspond to radii containing 0.5, 1, 3, 5, 10, 30, 50$\%$ of the total mass.}\label{fig:t_evol_param}
\end{center}
\end{figure*}

The NGC~6752 models reach core-collapsed at $\sim 8-9$~Gyr (Figure~\ref{fig:t_evol_param}), so naively we might expect that the single MSPs formed from the various dynamical channels would have enough time to experience exchange interactions and acquire binary companions by $\sim 12$~Gyr (the binary-single encounter timescale in a typical GC is $\lesssim 10^4$~yr). However, we still see many single MSPs observationally and in the models (Figure~\ref{fig:offset_6752} and Table~\ref{tab:clu_prop}) at the present day. This is because in the thermally balanced evolution of GCs (e.g., during core-collapse), the binary formation rate is equal to the binary ejection rate \citep[e.g.,][]{Antonini_Gieles_2020}, and the binary fractions of the heat source (WD and NS binaries in core-collapsed GCs) tend to a constant \citep[e.g.,][their Figure~12]{Kremer+2021WD}. We also observe this balance and the constant binary fractions in our models at late times ($\gtrsim 8$~Gyr), where the binary fraction of MSPs is $\sim 10\%$ (and the binary fraction of all NSs is $\sim 3\%$). The former is shown in the unvaried gaps between the number of `All' and `Single' systems in Figure~\ref{fig:t_evol_param} (bottom panels). In addition, among all the MSPs formed in TDEs and WD coalescences in the simulations, $\sim 30-40\%$ escaped the host clusters, and $\sim 30\%$ of the escapers are in binary.

\newpage
\section{Discussion}\label{sec:discuss}

We have shown that either efficient accretion during NS--main-sequence star TDEs, or direct formation of MSPs through merger-induced collapses of heavy WDs can boost the number of single MSPs in GCs. These dynamical processes allow the ratio of single-to-binary MSPs to match better with the observed ratio (eight for NGC~6752). Among the models in Table~\ref{tab:clu_prop}, model 6a and 6b have the largest $r_{ib}$, and thus are the most efficient at forming single MSPs compared to binary MSPs. Both models have $\sim150$ MSPs at the present day. This number is much larger than the current number of observed MSPs in NGC~6752. However, the detected number of MSPs in a cluster is affected by various selection biases from radio observations, and the true underlying population is unknown. X-ray observations may provide further constraints on the number of MSPs in a GC. A recent deep search using \textit{Chandra} has found $\sim50$ sources in NGC~6752, the origins of many already identified \citep{Cohn+2021}. This suggests that the number of MSPs in the cluster may be $\lesssim 50$ \citep[][and private communication with Craig Heinke]{Cohn+2021}. Therefore, many of the models listed would be overproducing MSPs if we assume very efficient mass transport onto the NSs during main-sequence star TDEs and all merger-induced collapses directly lead to MSPs.

It is possible that not all main-sequence TDEs efficiently spin up the participating NSs, as the detailed accretion mechanism is uncertain \citep[e.g.,][]{Kremer_nstde_2022}; or the NSs may rapidly spin down after the TDEs and power FRBs instead \citep{Li_Pen_2023}. It is also possible that not all super-Chandrasehkar WD--WD coalescences produce standard MSPs, especially if large magnetic fields can be generated during the merger and the subsequent collapse. Alternative outcomes include magnetars \citep[e.g.,][]{Usov_1992,Levan+2006}, Crab-like pulsars \citep[e.g.,][]{Kremer+2023}, or, in the case of detonation, type Ia supernovae \citep[e.g.,][]{Iben_Tutukov_1984,Webbink_1984,Maoz_Mannucci_2012,Pakmor+2012,Maoz+2014}. For example, single WDs in the Milky Way expected to have formed from WD mergers have observed magnetic fields of $\sim10^8$~G \citep[e.g.,][]{Ferrario2015,Caiazzo2021}, possibly created via dynamo processes in the hot, differentially rotating merger remnants \citep[e.g.,][]{Garcia-Berro+2012}. Flux conversion during the subsequent collapse to an NS would lead to an increase of order $(R_{\rm{WD}}/R_{\rm{NS}})^2$, suggesting field strengths $\sim10^{12}$~G for the final NS \citep[e.g.,][]{Levan+2006}. The relatively short spin-down timescales ($\lesssim 10^8\,$yr) of these NSs are inconsistent with standard MSPs with $B\lesssim10^9\,$G, but may be consistent with the four apparently young pulsars observed in several Milky Way GCs \citep{Boyles2011,Kremer+2023}. Such NSs formed via WD mergers may also power fast radio bursts \citep[FRBs; e.g.,][]{Margalit2019,Kremer+2021frb,Lu2022}, potentially connected with the repeating FRB observed in a GC in M81 \citep{Bhardwaj+2021,Kirsten+2022}. 

%\startlongtable
\begin{deluxetable}{c | c c c c}
\tabletypesize{\scriptsize}
\tablewidth{-1pt}
\setlength{\tabcolsep}{3pt}
\tablecaption{Rates of Super-Chandrasekhar WD--WD Coalescence} \label{tab:collrate}
\tablehead{\colhead{Models} & \colhead{Total} & \colhead{CO+CO} & \colhead{CO/ONe+ONe} & \colhead{Volumetric rate}\\
\colhead{} & \colhead{$10^{-8}\,\rm{yr^{-1}}$} & \colhead{$10^{-8}\,\rm{yr^{-1}}$} & \colhead{$10^{-8}\,\rm{yr^{-1}}$} & \colhead{$\rm{Gpc^{-3} yr^{-1}}$}
}
\startdata
a & $1.6$ & $0.88$ & $0.74$ & 37\\
b & $2.5$ & $1.4$ & $1.1$ & 58
\enddata
\tablecomments{The second column shows the total collision and merger rates per typical core-collapsed GC, and the third and fourth columns show the rates that include two CO WDs and at least one ONe WD per GC, respectively. The last column shows the estimated total volumetric rates at the local Universe assuming a GC number density of $2.31~\rm{Mpc^{-3}}$ and that all GCs are core-collapsed. All rates are calculated for the local Universe at $>9$~Gyr ($z\lesssim0.5$).}
\end{deluxetable}

Furthermore, the bottom right panel of Figure~\ref{fig:t_evol_param} and the bottom panel of Figure~\ref{fig:wdmass_distr} show that $\sim 30$ fast-spinning NSs can be formed at very early times from primordial double WD binaries. If these NSs were born directly as MSPs, even non-core-collapsed clusters could have a large number and, potentially fraction, of single MSPs. This indicates that either dynamical mass transfer between two heavy WDs does not conserve mass as is discussed in Section~\ref{subsec:wdtctde}, or many heavy WD collapse products are type Ia supernovae or highly magnetized, fast-spinning NSs that are FRB sources and will spin down quickly, becoming unobservable at the present day.

The volumetric rates of these super-Chandrasehkar WD--WD coalescences for typical core-collapsed GCs in the local Universe ($z\lesssim0.5$) are $\sim50~\rm{Gpc^{-3}\,yr^{-1}}$ (Table~\ref{tab:collrate}; implicitly assuming that all GCs in the local Universe are similar to NGC~6752). These rates may potentially be consistent with the production of M81-like FRB sources if we assume that all WD--WD collisions and mergers lead to an FRB source with a lifetime $\sim10^5$~year as proposed in \citet[][Eq.1]{Kremer+2021frb}. At the same time, NS accretion during TDEs would need to be efficient (Table~\ref{tab:clu_prop}) in order to explain the observed single MSPs. However, if all collisions or mergers involving two CO WDs lead to central carbon ignition and type Ia supernovae, the rate for forming FRB sources through the WD coalescence channel would be halved (Table~\ref{tab:collrate}). This would make it harder to explain FRB sources in GCs from this channel alone.

\section{Conclusions}\label{sec:conclu}
In this study, we have explored different formation channels for single MSPs in GCs using state-of-the-art Monte Carlo $N$-body simulations, focusing on a representative core-collapsed cluster NGC~6752. We ran 18 simulations whose present-day properties agree with the observed surface brightness profile and velocity dispersion profile of NGC~6752, and varied the dynamical processes for single MSP formation. In the dense stellar environments, single MSPs can form through binary-mediated close encounters which disrupt the original MSP binaries, spinning up of NSs through main-sequence star TDEs, direct formation of MSPs through merger-induced collapses of heavy WDs, or evaporation of the very low-mass, unstable MSP companion star. We have shown that binary disruption and evaporation of the companion stars alone are inadequate in explaining the observed single-to-binary MSP ratio. Instead, we have demonstrated that both main-sequence star TDEs and merger-induced collapses can significantly boost the number of single MSPs in GCs, and either one is necessary for better agreements with the observation, especially the ratio of single-to-binary MSPs. 

However, allowing both a high mass transport efficiency during TDEs and all super-Chandrasehkar WD--WD coalescences to directly collapse to MSPs may produce an overabundance of MSPs in core-collapsed clusters. Specifically, the collapses of heavy WD coalescence products to MSPs from primordial double WD binaries may also produce an overabundance of single MSPs in non-core-collapsed clusters. This overproduction can be alleviated if not all NSs are spun up efficiently during TDEs, and/or not all merger-induced collapses form MSPs. In particular, WD--WD collisions and mergers may produce fast-spinning NSs with moderate magnetic fields instead, which can emit FRBs similar to the one observed in the M81 GC \citep{Kremer+2021frb}. Future detections of additional FRBs in GCs of other nearby galaxies may provide further constraints on the potential connection between FRBs and NSs formed via WD coalescences in clusters \citep[e.g.,][]{Kremer+2023_detectfrb}.

\appendix
\restartappendixnumbering
\section{Uncertainties of White Dwarf Tidal Capture Interactions}\label{sec:append}

Low-mass WDs are represented by $n=1.5$ polytropes, while WDs with masses close to the Chandrasekhar limit can be modeled as $n=3$ polytropes \citep{Shapiro_Teukolsky_1983}. Although most WDs interacting in a core-collapsed GC have masses in between \citep[e.g.,][]{Kremer+2021WD}, we assume the WDs in our models are all $n=1.5$ polytropes for simplicity. Varying the polytropic index does not significantly affect the maximum pericenter distance for WD--WD tidal captures.

\begin{figure}[h!]
\begin{center}
\includegraphics[width=0.5\columnwidth]{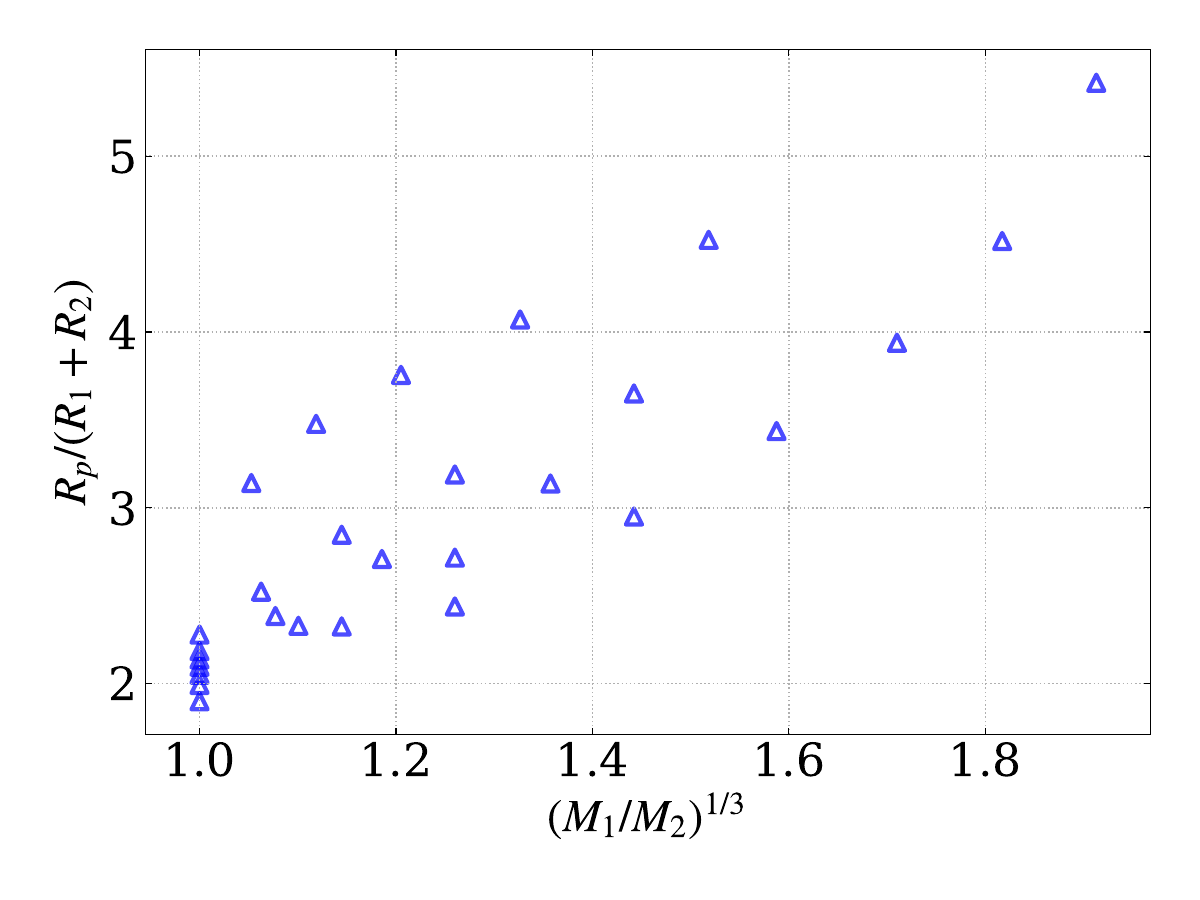}
\caption{Maximum pericenter distance for WD--WD tidal capture in the unit of the total WD radii as a function of the WD mass ratio. }\label{app:crosec_mration}
\end{center}
\end{figure}

Following \citet[][their Section~2.1, and references therein]{Ye_47tuc_2021}, we define the maximum tidal capture radius to be the pericenter distance during the first passage where the oscillation energy deposited into both WDs exceeds that of their initial total kinetic energy at infinity. We show the maximum tidal capture radii for different WD masses (0.2-1.4$~\msun$) and mass ratios with a constant velocity dispersion of $10~\rm{km\,s^{-1}}$ in Figure~\ref{app:crosec_mration}. For mass ratios close to unity, the maximum tidal capture radius is about twice the sum of the WD radii. Thus we simply use $2(R_1+R_2)$, where $R_1$ and $R_2$ are the radii of the WDs, as the maximum tidal capture radius in our simulations since most interactions in GCs are between objects with similar masses due to mass segregation.

\newpage
\begin{acknowledgements}
    We thank Chris Thompson and Ue-Li Pen for helpful discussions, and the anonymous referee for helpful comments and suggestions. C.S.Y. acknowledges support from the Natural Sciences and Engineering Research Council of Canada (NSERC) DIS-2022-568580. Support for KK was provided by NASA through the NASA Hubble Fellowship grant HST-HF2-51510 awarded by the Space Telescope Science Institute, which is operated by the Association of Universities for Research in Astronomy, Inc., for NASA, under contract NAS5-26555. SMR is a CIFAR Fellow and is supported by the NSF Physics Frontiers Center award 2020265. F. A. R. acknowledges support from NSF grant AST-2108624 and NASA ATP grant 80NSSC22K0722. This research was supported in part through the computational resources and staff contributions provided for the Quest high performance computing facility at Northwestern University, which is jointly supported by the Office of the Provost, the Office for Research, and Northwestern University Information Technology. This work was performed in part at the Aspen Center for Physics, which is supported by National Science Foundation grant PHY-2210452, and was partially supported by a grant from the Simons Foundation. The National Radio Astronomy Observatory is a facility of the National Science Foundation operated under cooperative agreement by Associated Universities, Inc.
\end{acknowledgements}

\software{\texttt{CMC} \citep{Joshi_2000,Joshi_2001,Fregeau_2003, fregeau2007monte, Chatterjee_2010,Chatterjee_2013b,Umbreit_2012,Morscher+2015,Rodriguez+2016million,CMC1}, \texttt{Fewbody} \citep{fregeau2004stellar}, \texttt{COSMIC} \citep{cosmic}}

\bibliography{iso_msp}

\begin{thebibliography}{}
\expandafter\ifx\csname natexlab\endcsname\relax\def\natexlab#1{#1}\fi
\providecommand{\url}[1]{\href{#1}{#1}}
\providecommand{\dodoi}[1]{doi:~\href{http://doi.org/#1}{\nolinkurl{#1}}}
\providecommand{\doeprint}[1]{\href{http://ascl.net/#1}{\nolinkurl{http://ascl.net/#1}}}
\providecommand{\doarXiv}[1]{\href{https://arxiv.org/abs/#1}{\nolinkurl{https://arxiv.org/abs/#1}}}

\bibitem[{{Abbott} {et~al.}(2020){Abbott}, {Abbott}, {Abraham}, {Acernese},
  {Ackley}, {Adams}, {Adhikari}, {Adya}, {Affeldt}, {Agathos}, {Agatsuma},
  {Aggarwal}, {Aguiar}, {Aich}, {Aiello}, {Ain}, {Ajith}, {Akcay}, {Allen},
  {Allocca}, {Altin}, {Amato}, {Anand}, {Ananyeva}, {Anderson}, {Anderson},
  {Angelova}, {Ansoldi}, {Antier}, {Appert}, {Arai}, {Araya}, {Areeda},
  {Ar{\`e}ne}, {Arnaud}, {Aronson}, {Arun}, {Asali}, {Ascenzi}, {Ashton},
  {Aston}, {Astone}, {Aubin}, {Aufmuth}, {AultONeal}, {Austin}, {Avendano},
  {Babak}, {Bacon}, {Badaracco}, {Bader}, {Bae}, {Baer}, {Baird}, {Baldaccini},
  {Ballardin}, {Ballmer}, {Bals}, {Balsamo}, {Baltus}, {Banagiri}, {Bankar},
  {Bankar}, {Barayoga}, {Barbieri}, {Barish}, {Barker}, {Barkett}, {Barneo},
  {Barone}, {Barr}, {Barsotti}, {Barsuglia}, {Barta}, {Bartlett}, {Bartos},
  {Bassiri}, {Basti}, {Bawaj}, {Bayley}, {Bazzan}, {B{\'e}csy}, {Bejger},
  {Belahcene}, {Bell}, {Beniwal}, {Benjamin}, {Benkel}, {Bentley}, {Bergamin},
  {Berger}, {Bergmann}, {Bernuzzi}, {Berry}, {Bersanetti}, {Bertolini},
  {Betzwieser}, {Bhandare}, {Bhandari}, {Bidler}, {Biggs}, {Bilenko},
  {Billingsley}, {Birney}, {Birnholtz}, {Biscans}, {Bischi}, {Biscoveanu},
  {Bisht}, {Bissenbayeva}, {Bitossi}, {Bizouard}, {Blackburn}, {Blackman},
  {Blair}, {Blair}, {Blair}, {Bobba}, {Bode}, {Boer}, {Boetzel}, {Bogaert},
  {Bondu}, {Bonilla}, {Bonnand}, {Booker}, {Boom}, {Bork}, {Boschi}, {Bose},
  {Bossilkov}, {Bosveld}, {Bouffanais}, {Bozzi}, {Bradaschia}, {Brady},
  {Bramley}, {Branchesi}, {Brau}, {Breschi}, {Briant}, {Briggs}, {Brighenti},
  {Brillet}, {Brinkmann}, {Brito}, {Brockill}, {Brooks}, {Brooks}, {Brown},
  {Brunett}, {Bruno}, {Bruntz}, {Buikema}, {Bulik}, {Bulten}, {Buonanno},
  {Buskulic}, {Byer}, {Cabero}, {Cadonati}, {Cagnoli}, {Cahillane}, {Bustillo},
  {Callaghan}, {Callister}, {Calloni}, {Camp}, {Canepa}, {Cannon}, {Cao},
  {Cao}, {Carapella}, {Carbognani}, {Caride}, {Carney}, {Carullo}, {Diaz},
  {Casentini}, {Casta{\~n}eda}, {Caudill}, {Cavagli{\`a}}, {Cavalier},
  {Cavalieri}, {Cella}, {Cerd{\'a}-Dur{\'a}n}, {Cesarini}, {Chaibi},
  {Chakravarti}, {Chan}, {Chan}, {Chao}, {Charlton}, {Chase},
  {Chassande-Mottin}, {Chatterjee}, {Chaturvedi}, {Chatziioannou}, {Chen},
  {Chen}, {Chen}, {Cheng}, {Cheong}, {Chia}, {Chiadini}, {Chierici},
  {Chincarini}, {Chiummo}, {Cho}, {Cho}, {Cho}, {Christensen}, {Chu}, {Chua},
  {Chung}, {Chung}, {Ciani}, {Ciecielag}, {Cie{\'s}lar}, {Ciobanu}, {Ciolfi},
  {Cipriano}, {Cirone}, {Clara}, {Clark}, {Clearwater}, {Clesse}, {Cleva},
  {Coccia}, {Cohadon}, {Cohen}, {Colleoni}, {Collette}, {Collins}, {Colpi},
  {Constancio}, {Conti}, {Cooper}, {Corban}, {Corbitt}, {Cordero-Carri{\'o}n},
  {Corezzi}, {Corley}, {Cornish}, {Corre}, {Corsi}, {Cortese}, {Costa},
  {Cotesta}, {Coughlin}, {Coughlin}, {Coulon}, {Countryman}, {Couvares},
  {Covas}, {Coward}, {Cowart}, {Coyne}, {Coyne}, {Creighton}, {Creighton},
  {Cripe}, {Croquette}, {Crowder}, {Cudell}, {Cullen}, {Cumming}, {Cummings},
  {Cunningham}, {Cuoco}, {Curylo}, {Canton}, {D{\'a}lya}, {Dana},
  {Daneshgaran-Bajastani}, {D'Angelo}, {Danilishin}, {D'Antonio}, {Danzmann},
  {Darsow-Fromm}, {Dasgupta}, {Datrier}, {Dattilo}, {Dave}, {Davier}, {Davies},
  {Davis}, {Daw}, {DeBra}, {Deenadayalan}, {Degallaix}, {De Laurentis},
  {Del{\'e}glise}, {Delfavero}, {De Lillo}, {Del Pozzo}, {DeMarchi},
  {D'Emilio}, {Demos}, {Dent}, {De Pietri}, {De Rosa}, {De Rossi}, {DeSalvo},
  {de Varona}, {Dhurandhar}, {D{\'\i}az}, {Diaz-Ortiz}, {Dietrich}, {Di Fiore},
  {Di Fronzo}, {Di Giorgio}, {Di Giovanni}, {Di Giovanni}, {Di Girolamo}, {Di
  Lieto}, {Ding}, {Di Pace}, {Di Palma}, {Di Renzo}, {Divakarla}, {Dmitriev},
  {Doctor}, {Donovan}, {Dooley}, {Doravari}, {Dorrington}, {Downes}, {Drago},
  {Driggers}, {Du}, {Ducoin}, {Dupej}, {Durante}, {D'Urso}, {Dwyer}, {Easter},
  {Eddolls}, {Edelman}, {Edo}, {Edy}, {Effler}, {Ehrens}, {Eichholz},
  {Eikenberry}, {Eisenmann}, {Eisenstein}, {Ejlli}, {Errico}, {Essick},
  {Estelles}, {Estevez}, {Etienne}, {Etzel}, {Evans}, {Evans}, {Ewing},
  {Fafone}, {Fairhurst}, {Fan}, {Farinon}, {Farr}, {Farr}, {Fauchon-Jones},
  {Favata}, {Fays}, {Fazio}, {Feicht}, {Fejer}, {Feng}, {Fenyvesi}, {Ferguson},
  {Fernandez-Galiana}, {Ferrante}, {Ferreira}, {Ferreira}, {Fidecaro}, {Fiori},
  {Fiorucci}, {Fishbach}, {Fisher}, {Fittipaldi}, {Fitz-Axen}, {Fiumara},
  {Flaminio}, {Floden}, {Flynn}, {Fong}, {Font}, {Forsyth}, {Fournier},
  {Frasca}, {Frasconi}, {Frei}, {Freise}, {Frey}, {Frey}, {Fritschel},
  {Frolov}, {Fronz{\`e}}, {Fulda}, {Fyffe}, {Gabbard}, {Gadre}, {Gaebel},
  {Gair}, {Galaudage}, {Ganapathy}, {Ganguly}, {Gaonkar},
  {Garc{\'\i}a-Quir{\'o}s}, {Garufi}, {Gateley}, {Gaudio}, {Gayathri}, {Gemme},
  {Genin}, {Gennai}, {George}, {George}, {Gergely}, {Ghonge}, {Ghosh}, {Ghosh},
  {Ghosh}, {Giacomazzo}, {Giaime}, {Giardina}, {Gibson}, {Gier}, {Gill},
  {Glanzer}, {Gniesmer}, {Godwin}, {Goetz}, {Goetz}, {Gohlke}, {Goncharov},
  {Gonz{\'a}lez}, {Gopakumar}, {Gossan}, {Gosselin}, {Gouaty}, {Grace},
  {Grado}, {Granata}, {Grant}, {Gras}, {Grassia}, {Gray}, {Gray}, {Greco},
  {Green}, {Green}, {Gretarsson}, {Griggs}, {Grignani}, {Grimaldi}, {Grimm},
  {Grote}, {Grunewald}, {Gruning}, {Guidi}, {Guimaraes}, {Guix{\'e}}, {Gulati},
  {Guo}, {Gupta}, {Gupta}, {Gupta}, {Gustafson}, {Gustafson}, {Haegel},
  {Halim}, {Hall}, {Hamilton}, {Hammond}, {Haney}, {Hanke}, {Hanks}, {Hanna},
  {Hannam}, {Hannuksela}, {Hansen}, {Hanson}, {Harder}, {Hardwick}, {Haris},
  {Harms}, {Harry}, {Harry}, {Hasskew}, {Haster}, {Haughian}, {Hayes}, {Healy},
  {Heidmann}, {Heintze}, {Heinze}, {Heitmann}, {Hellman}, {Hello}, {Hemming},
  {Hendry}, {Heng}, {Hennes}, {Hennig}, {Heurs}, {Hild}, {Hinderer}, {Hoback},
  {Hochheim}, {Hofgard}, {Hofman}, {Holgado}, {Holland}, {Holt}, {Holz},
  {Hopkins}, {Horst}, {Hough}, {Howell}, {Hoy}, {Huang}, {H{\"u}bner},
  {Huerta}, {Huet}, {Hughey}, {Hui}, {Husa}, {Huttner}, {Huxford},
  {Huynh-Dinh}, {Idzkowski}, {Iess}, {Inchauspe}, {Ingram}, {Intini}, {Isac},
  {Isi}, {Iyer}, {Jacqmin}, {Jadhav}, {Jadhav}, {James}, {Jani}, {Janthalur},
  {Jaranowski}, {Jariwala}, {Jaume}, {Jenkins}, {Jiang}, {Johns},
  {Johnson-McDaniel}, {Jones}, {Jones}, {Jones}, {Jones}, {Jones}, {Jonker},
  {Ju}, {Junker}, {Kalaghatgi}, {Kalogera}, {Kamai}, {Kandhasamy}, {Kang},
  {Kanner}, {Kapadia}, {Karki}, {Kashyap}, {Kasprzack}, {Kastaun},
  {Katsanevas}, {Katsavounidis}, {Katzman}, {Kaufer}, {Kawabe},
  {K{\'e}f{\'e}lian}, {Keitel}, {Keivani}, {Kennedy}, {Key}, {Khadka},
  {Khalili}, {Khan}, {Khan}, {Khan}, {Khazanov}, {Khetan}, {Khursheed},
  {Kijbunchoo}, {Kim}, {Kim}, {Kim}, {Kim}, {Kim}, {Kim}, {Kim}, {Kimball},
  {King}, {Kinley-Hanlon}, {Kirchhoff}, {Kissel}, {Kleybolte}, {Klimenko},
  {Knowles}, {Knyazev}, {Koch}, {Koehlenbeck}, {Koekoek}, {Koley},
  {Kondrashov}, {Kontos}, {Koper}, {Korobko}, {Korth}, {Kovalam}, {Kozak},
  {Kringel}, {Krishnendu}, {Kr{\'o}lak}, {Krupinski}, {Kuehn}, {Kumar},
  {Kumar}, {Kumar}, {Kumar}, {Kumar}, {Kuo}, {Kutynia}, {Lackey}, {Laghi},
  {Lalande}, {Lam}, {Lamberts}, {Landry}, {Landry}, {Lane}, {Lang}, {Lange},
  {Lantz}, {Lanza}, {La Rosa}, {Lartaux-Vollard}, {Lasky}, {Laxen},
  {Lazzarini}, {Lazzaro}, {Leaci}, {Leavey}, {Lecoeuche}, {Lee}, {Lee}, {Lee},
  {Lee}, {Lee}, {Lehmann}, {Leroy}, {Letendre}, {Levin}, {Li}, {Li}, {li},
  {Li}, {Li}, {Linde}, {Linker}, {Linley}, {Littenberg}, {Liu}, {Liu},
  {Llorens-Monteagudo}, {Lo}, {Lockwood}, {London}, {Longo}, {Lorenzini},
  {Loriette}, {Lormand}, {Losurdo}, {Lough}, {Lousto}, {Lovelace}, {L{\"u}ck},
  {Lumaca}, {Lundgren}, {Ma}, {Macas}, {Macfoy}, {MacInnis}, {Macleod},
  {MacMillan}, {Macquet}, {Hernandez}, {Maga{\~n}a-Sandoval}, {Magee},
  {Majorana}, {Maksimovic}, {Malik}, {Man}, {Mandic}, {Mangano}, {Mansell},
  {Manske}, {Mantovani}, {Mapelli}, {Marchesoni}, {Marion}, {M{\'a}rka},
  {M{\'a}rka}, {Markakis}, {Markosyan}, {Markowitz}, {Maros}, {Marquina},
  {Marsat}, {Martelli}, {Martin}, {Martin}, {Martinez}, {Martynov},
  {Masalehdan}, {Mason}, {Massera}, {Masserot}, {Massinger}, {Masso-Reid},
  {Mastrogiovanni}, {Matas}, {Matichard}, {Mavalvala}, {Maynard}, {McCann},
  {McCarthy}, {McClelland}, {McCormick}, {McCuller}, {McGuire}, {McIsaac},
  {McIver}, {McManus}, {McRae}, {McWilliams}, {Meacher}, {Meadors}, {Mehmet},
  {Mehta}, {Villa}, {Melatos}, {Mendell}, {Mercer}, {Mereni}, {Merfeld},
  {Merilh}, {Merritt}, {Merzougui}, {Meshkov}, {Messenger}, {Messick},
  {Metzdorff}, {Meyers}, {Meylahn}, {Mhaske}, {Miani}, {Miao}, {Michaloliakos},
  {Michel}, {Middleton}, {Milano}, {Miller}, {Millhouse}, {Mills}, {Milotti},
  {Milovich-Goff}, {Minazzoli}, {Minenkov}, {Mishkin}, {Mishra}, {Mistry},
  {Mitra}, {Mitrofanov}, {Mitselmakher}, {Mittleman}, {Mo}, {Mogushi},
  {Mohapatra}, {Mohite}, {Molina-Ruiz}, {Mondin}, {Montani}, {Moore}, {Moraru},
  {Morawski}, {Moreno}, {Morisaki}, {Mours}, {Mow-Lowry}, {Mozzon},
  {Muciaccia}, {Mukherjee}, {Mukherjee}, {Mukherjee}, {Mukherjee}, {Mukund},
  {Mullavey}, {Munch}, {Mu{\~n}iz}, {Murray}, {Nagar}, {Nardecchia},
  {Naticchioni}, {Nayak}, {Neil}, {Neilson}, {Nelemans}, {Nelson}, {Nery},
  {Neunzert}, {Ng}, {Ng}, {Nguyen}, {Nguyen}, {Nichols}, {Nichols}, {Nissanke},
  {Nocera}, {Noh}, {North}, {Nothard}, {Nuttall}, {Oberling}, {O'Brien},
  {Oganesyan}, {Ogin}, {Oh}, {Oh}, {Ohme}, {Ohta}, {Okada}, {Oliver},
  {Olivetto}, {Oppermann}, {Oram}, {O'Reilly}, {Ormiston}, {Ortega},
  {O'Shaughnessy}, {Ossokine}, {Osthelder}, {Ottaway}, {Overmier}, {Owen},
  {Pace}, {Pagano}, {Page}, {Pagliaroli}, {Pai}, {Pai}, {Palamos}, {Palashov},
  {Palomba}, {Pan}, {Panda}, {Pang}, {Pankow}, {Pannarale}, {Pant}, {Paoletti},
  {Paoli}, {Parida}, {Parker}, {Pascucci}, {Pasqualetti}, {Passaquieti},
  {Passuello}, {Patricelli}, {Payne}, {Pearlstone}, {Pechsiri}, {Pedersen},
  {Pedraza}, {Pele}, {Penn}, {Perego}, {Perez}, {P{\'e}rigois}, {Perreca},
  {Perri{\`e}s}, {Petermann}, {Pfeiffer}, {Phelps}, {Phukon}, {Piccinni},
  {Pichot}, {Piendibene}, {Piergiovanni}, {Pierro}, {Pillant}, {Pinard},
  {Pinto}, {Piotrzkowski}, {Pirello}, {Pitkin}, {Plastino}, {Poggiani}, {Pong},
  {Ponrathnam}, {Popolizio}, {Porter}, {Powell}, {Prajapati}, {Prasai},
  {Prasanna}, {Pratten}, {Prestegard}, {Principe}, {Prodi}, {Prokhorov},
  {Punturo}, {Puppo}, {P{\"u}rrer}, {Qi}, {Quetschke}, {Quinonez}, {Raab},
  {Raaijmakers}, {Radkins}, {Radulesco}, {Raffai}, {Rafferty}, {Raja}, {Rajan},
  {Rajbhandari}, {Rakhmanov}, {Ramirez}, {Ramos-Buades}, {Rana}, {Rao},
  {Rapagnani}, {Raymond}, {Razzano}, {Read}, {Regimbau}, {Rei}, {Reid},
  {Reitze}, {Rettegno}, {Ricci}, {Richardson}, {Richardson}, {Ricker},
  {Riemenschneider}, {Riles}, {Rizzo}, {Robertson}, {Robinet}, {Rocchi},
  {Rodriguez-Soto}, {Rolland}, {Rollins}, {Roma}, {Romanelli}, {Romano},
  {Romel}, {Romero-Shaw}, {Romie}, {Rose}, {Rose}, {Rose}, {Rosi{\'n}ska},
  {Rosofsky}, {Ross}, {Rowan}, {Rowlinson}, {Roy}, {Roy}, {Roy}, {Ruggi},
  {Rutins}, {Ryan}, {Sachdev}, {Sadecki}, {Sakellariadou}, {Salafia},
  {Salconi}, {Saleem}, {Salemi}, {Samajdar}, {Sanchez}, {Sanchez},
  {Sanchis-Gual}, {Sanders}, {Santiago}, {Santos}, {Sarin}, {Sassolas},
  {Sathyaprakash}, {Sauter}, {Savage}, {Savant}, {Sawant}, {Sayah}, {Schaetzl},
  {Schale}, {Scheel}, {Scheuer}, {Schmidt}, {Schnabel}, {Schofield},
  {Sch{\"o}nbeck}, {Schreiber}, {Schulte}, {Schutz}, {Schwarm}, {Schwartz},
  {Scott}, {Scott}, {Seidel}, {Sellers}, {Sengupta}, {Sennett}, {Sentenac},
  {Sequino}, {Sergeev}, {Setyawati}, {Shaddock}, {Shaffer}, {Shahriar},
  {Sharma}, {Sharma}, {Shawhan}, {Shen}, {Shikauchi}, {Shink}, {Shoemaker},
  {Shoemaker}, {Shukla}, {ShyamSundar}, {Siellez}, {Sieniawska}, {Sigg},
  {Singer}, {Singh}, {Singh}, {Singha}, {Singhal}, {Sintes}, {Sipala},
  {Skliris}, {Slagmolen}, {Slaven-Blair}, {Smetana}, {Smith}, {Smith},
  {Somala}, {Son}, {Soni}, {Sorazu}, {Sordini}, {Sorrentino}, {Souradeep},
  {Sowell}, {Spencer}, {Spera}, {Srivastava}, {Srivastava}, {Staats},
  {Stachie}, {Standke}, {Steer}, {Steinhoff}, {Steinke}, {Steinlechner},
  {Steinlechner}, {Steinmeyer}, {Stevenson}, {Stocks}, {Stops}, {Stover},
  {Strain}, {Stratta}, {Strunk}, {Sturani}, {Stuver}, {Sudhagar}, {Sudhir},
  {Summerscales}, {Sun}, {Sunil}, {Sur}, {Suresh}, {Sutton}, {Swinkels},
  {Szczepa{\'n}czyk}, {Tacca}, {Tait}, {Talbot}, {Tanasijczuk}, {Tanner},
  {Tao}, {T{\'a}pai}, {Tapia}, {San Martin}, {Tasson}, {Taylor}, {Tenorio},
  {Terkowski}, {Thirugnanasambandam}, {Thomas}, {Thomas}, {Thompson},
  {Thondapu}, {Thorne}, {Thrane}, {Tinsman}, {Saravanan}, {Tiwari}, {Tiwari},
  {Tiwari}, {Toland}, {Tonelli}, {Tornasi}, {Torres-Forn{\'e}}, {Torrie},
  {Tosta e Melo}, {T{\"o}yr{\"a}}, {Trail}, {Travasso}, {Traylor}, {Tringali},
  {Tripathee}, {Trovato}, {Trudeau}, {Tsang}, {Tse}, {Tso}, {Tsukada}, {Tsuna},
  {Tsutsui}, {Turconi}, {Ubhi}, {Ueno}, {Ugolini}, {Unnikrishnan}, {Urban},
  {Usman}, {Utina}, {Vahlbruch}, {Vajente}, {Valdes}, {Valentini}, {van Bakel},
  {van Beuzekom}, {van den Brand}, {Van Den Broeck}, {Vander-Hyde}, {van der
  Schaaf}, {Van Heijningen}, {van Veggel}, {Vardaro}, {Varma}, {Vass},
  {Vas{\'u}th}, {Vecchio}, {Vedovato}, {Veitch}, {Veitch}, {Venkateswara},
  {Venugopalan}, {Verkindt}, {Veske}, {Vetrano}, {Vicer{\'e}}, {Viets},
  {Vinciguerra}, {Vine}, {Vinet}, {Vitale}, {Vivanco}, {Vo}, {Vocca},
  {Vorvick}, {Vyatchanin}, {Wade}, {Wade}, {Wade}, {Walet}, {Walker},
  {Wallace}, {Wallace}, {Walsh}, {Wang}, {Wang}, {Wang}, {Ward}, {Warden},
  {Warner}, {Was}, {Watchi}, {Weaver}, {Wei}, {Weinert}, {Weinstein}, {Weiss},
  {Wellmann}, {Wen}, {We{\ss}els}, {Westhouse}, {Wette}, {Whelan}, {Whiting},
  {Whittle}, {Wilken}, {Williams}, {Willis}, {Willke}, {Winkler}, {Wipf},
  {Wittel}, {Woan}, {Woehler}, {Wofford}, {Wong}, {Wright}, {Wu}, {Wysocki},
  {Xiao}, {Yamamoto}, {Yang}, {Yang}, {Yang}, {Yap}, {Yazback}, {Yeeles}, {Yu},
  {Yu}, {Yuen}, {Zadro{\.z}ny}, {Zadro{\.z}ny}, {Zanolin}, {Zelenova},
  {Zendri}, {Zevin}, {Zhang}, {Zhang}, {Zhang}, {Zhao}, {Zhao}, {Zhou}, {Zhou},
  {Zhu}, {Zimmerman}, {Zucker}, {Zweizig}, {LIGO Scientific Collaboration}, \&
  {Virgo Collaboration}}]{gw190814}
{Abbott}, R., {Abbott}, T.~D., {Abraham}, S., {et~al.} 2020, \apjl, 896, L44,
  \dodoi{10.3847/2041-8213/ab960f}

\bibitem[{{Alpar} {et~al.}(1982){Alpar}, {Cheng}, {Ruderman}, \&
  {Shaham}}]{Alpar+1982}
{Alpar}, M.~A., {Cheng}, A.~F., {Ruderman}, M.~A., \& {Shaham}, J. 1982, \nat,
  300, 728, \dodoi{10.1038/300728a0}

\bibitem[{Amaro-Seoane \& Chen(2016)}]{amaro2016relativistic}
Amaro-Seoane, P., \& Chen, X. 2016, \mnras, 458, 3075

\bibitem[{{Andersen} \& {Ransom}(2018)}]{Andersen_Ransom_2018}
{Andersen}, B.~C., \& {Ransom}, S.~M. 2018, \apjl, 863, L13,
  \dodoi{10.3847/2041-8213/aad59f}

\bibitem[{Antognini {et~al.}(2014)Antognini, Shappee, Thompson, \&
  Amaro-Seoane}]{antognini2014rapid}
Antognini, J.~M., Shappee, B.~J., Thompson, T.~A., \& Amaro-Seoane, P. 2014,
  \mnras, 439, 1079

\bibitem[{{Antonini} \& {Gieles}(2020)}]{Antonini_Gieles_2020}
{Antonini}, F., \& {Gieles}, M. 2020, \mnras, 492, 2936,
  \dodoi{10.1093/mnras/stz3584}

\bibitem[{{Armitage} \& {Livio}(2000)}]{Armitage_Livio_2000}
{Armitage}, P.~J., \& {Livio}, M. 2000, \apj, 532, 540, \dodoi{10.1086/308548}

\bibitem[{{Bailyn} {et~al.}(1998){Bailyn}, {Jain}, {Coppi}, \&
  {Orosz}}]{Bailyn+1998}
{Bailyn}, C.~D., {Jain}, R.~K., {Coppi}, P., \& {Orosz}, J.~A. 1998, \apj, 499,
  367, \dodoi{10.1086/305614}

\bibitem[{{Bassa} {et~al.}(2006){Bassa}, {van Kerkwijk}, {Koester}, \&
  {Verbunt}}]{Bassa+2006}
{Bassa}, C.~G., {van Kerkwijk}, M.~H., {Koester}, D., \& {Verbunt}, F. 2006,
  \aap, 456, 295, \dodoi{10.1051/0004-6361:20065181}

\bibitem[{{Baumgardt}(2017)}]{Baumgardt_2017}
{Baumgardt}, H. 2017, \mnras, 464, 2174, \dodoi{10.1093/mnras/stw2488}

\bibitem[{{Baumgardt} \& {Hilker}(2018)}]{Baumgardt_Hilker2018}
{Baumgardt}, H., \& {Hilker}, M. 2018, \mnras, 478, 1520,
  \dodoi{10.1093/mnras/sty1057}

\bibitem[{{Baumgardt} \& {Vasiliev}(2021)}]{Baumgardt_Vasiliev_2021}
{Baumgardt}, H., \& {Vasiliev}, E. 2021, \mnras, 505, 5957,
  \dodoi{10.1093/mnras/stab1474}

\bibitem[{{Beccari} {et~al.}(2006){Beccari}, {Ferraro}, {Possenti}, {Valenti},
  {Origlia}, \& {Rood}}]{Beccari+2006}
{Beccari}, G., {Ferraro}, F.~R., {Possenti}, A., {et~al.} 2006, \aj, 131, 2551,
  \dodoi{10.1086/500643}

\bibitem[{{Bedin} {et~al.}(2023){Bedin}, {Salaris}, {Anderson}, {Scalco},
  {Nardiello}, {Vesperini}, {Richer}, {Burgasser}, {Griggio}, {Gerasimov},
  {Apai}, {Bellini}, {Libralato}, {Bergeron}, {Rich}, \&
  {Grazian}}]{Bedin+2023}
{Bedin}, L.~R., {Salaris}, M., {Anderson}, J., {et~al.} 2023, \mnras, 518,
  3722, \dodoi{10.1093/mnras/stac3219}

\bibitem[{{B{\'e}gin}(2006)}]{Begin_2006}
{B{\'e}gin}, S. 2006, Master's thesis, University of British Columbia, Canada

\bibitem[{{Belczynski} {et~al.}(2008){Belczynski}, {Kalogera}, {Rasio}, {Taam},
  {Zezas}, {Bulik}, {Maccarone}, \& {Ivanova}}]{Belczynski+2008}
{Belczynski}, K., {Kalogera}, V., {Rasio}, F.~A., {et~al.} 2008, \apjs, 174,
  223, \dodoi{10.1086/521026}

\bibitem[{{Belczynski} {et~al.}(2010){Belczynski}, {Lorimer}, {Ridley}, \&
  {Curran}}]{Belczynski+2010}
{Belczynski}, K., {Lorimer}, D.~R., {Ridley}, J.~P., \& {Curran}, S.~J. 2010,
  \mnras, 407, 1245, \dodoi{10.1111/j.1365-2966.2010.16970.x}

\bibitem[{{Bhardwaj} {et~al.}(2021){Bhardwaj}, {Gaensler}, {Kaspi},
  {Landecker}, {Mckinven}, {Michilli}, {Pleunis}, {Tendulkar}, {Andersen},
  {Boyle}, {Cassanelli}, {Chawla}, {Cook}, {Dobbs}, {Fonseca}, {Kaczmarek},
  {Leung}, {Masui}, {Mnchmeyer}, {Ng}, {Rafiei-Ravandi}, {Scholz}, {Shin},
  {Smith}, {Stairs}, \& {Zwaniga}}]{Bhardwaj+2021}
{Bhardwaj}, M., {Gaensler}, B.~M., {Kaspi}, V.~M., {et~al.} 2021, \apjl, 910,
  L18, \dodoi{10.3847/2041-8213/abeaa6}

\bibitem[{{Bhattacharya} \& {van den
  Heuvel}(1991)}]{Bhattacharya_vandenHeuvel_1991}
{Bhattacharya}, D., \& {van den Heuvel}, E.~P.~J. 1991, \physrep, 203, 1,
  \dodoi{10.1016/0370-1573(91)90064-S}

\bibitem[{{Bildsten}(1998)}]{Bilsten1998}
{Bildsten}, L. 1998, in NATO Advanced Study Institute (ASI) Series C, Vol. 515,
  The Many Faces of Neutron Stars., ed. R.~{Buccheri}, J.~{van Paradijs}, \&
  A.~{Alpar}, 419, \dodoi{10.48550/arXiv.astro-ph/9709094}

\bibitem[{{Bildsten}(2002)}]{Bildsten_2002}
{Bildsten}, L. 2002, \apjl, 577, L27, \dodoi{10.1086/344085}

\bibitem[{{Blandford} \& {Begelman}(1999)}]{Blandford_Begelman_1999}
{Blandford}, R.~D., \& {Begelman}, M.~C. 1999, \mnras, 303, L1,
  \dodoi{10.1046/j.1365-8711.1999.02358.x}

\bibitem[{{Boyles} {et~al.}(2011){Boyles}, {Lorimer}, {Turk}, {Mnatsakanov},
  {Lynch}, {Ransom}, {Freire}, \& {Belczynski}}]{Boyles2011}
{Boyles}, J., {Lorimer}, D.~R., {Turk}, P.~J., {et~al.} 2011, \apj, 742, 51,
  \dodoi{10.1088/0004-637X/742/1/51}

\bibitem[{{Breivik} {et~al.}(2020){Breivik}, {Coughlin}, {Zevin}, {Rodriguez},
  {Kremer}, {Ye}, {Andrews}, {Kurkowski}, {Digman}, {Larson}, \&
  {Rasio}}]{cosmic}
{Breivik}, K., {Coughlin}, S., {Zevin}, M., {et~al.} 2020, \apj, 898, 71,
  \dodoi{10.3847/1538-4357/ab9d85}

\bibitem[{{Buonanno} {et~al.}(1986){Buonanno}, {Corsi}, {Iannicola}, \& {Fusi
  Pecci}}]{Buonanno+1986}
{Buonanno}, R., {Corsi}, C.~E., {Iannicola}, G., \& {Fusi Pecci}, F. 1986,
  \aap, 159, 189

\bibitem[{{Caiazzo} {et~al.}(2021){Caiazzo}, {Burdge}, {Fuller}, {Heyl},
  {Kulkarni}, {Prince}, {Richer}, {Schwab}, {Andreoni}, {Bellm}, {Drake},
  {Duev}, {Graham}, {Helou}, {Mahabal}, {Masci}, {Smith}, \&
  {Soumagnac}}]{Caiazzo2021}
{Caiazzo}, I., {Burdge}, K.~B., {Fuller}, J., {et~al.} 2021, \nat, 595, 39,
  \dodoi{10.1038/s41586-021-03615-y}

\bibitem[{{Camilo} {et~al.}(1993){Camilo}, {Nice}, \& {Taylor}}]{Camilo+1993}
{Camilo}, F., {Nice}, D.~J., \& {Taylor}, J.~H. 1993, \apjl, 412, L37,
  \dodoi{10.1086/186934}

\bibitem[{{Camilo} \& {Rasio}(2005)}]{Camilo_Rasio_2005}
{Camilo}, F., \& {Rasio}, F.~A. 2005, in Astronomical Society of the Pacific
  Conference Series, Vol. 328, Binary Radio Pulsars, ed. F.~A. {Rasio} \& I.~H.
  {Stairs}, 147, \dodoi{10.48550/arXiv.astro-ph/0501226}

\bibitem[{Chatterjee {et~al.}(2010)Chatterjee, Fregeau, Umbreit, \&
  Rasio}]{Chatterjee_2010}
Chatterjee, S., Fregeau, J.~M., Umbreit, S., \& Rasio, F.~A. 2010, \apj, 719,
  915.
\newblock \url{http://dx.doi.org/10.1088/0004-637X/719/1/915}

\bibitem[{Chatterjee {et~al.}(2013)Chatterjee, Umbreit, Fregeau, \&
  Rasio}]{Chatterjee_2013b}
Chatterjee, S., Umbreit, S., Fregeau, J.~M., \& Rasio, F.~A. 2013, \mnras, 429,
  2881.
\newblock \url{http://dx.doi.org/10.1093/mnras/sts464}

\bibitem[{{Cohn} {et~al.}(2021){Cohn}, {Lugger}, {Zhao}, {Tudor}, {Heinke},
  {Cool}, {Anderson}, {Strader}, \& {Miller-Jones}}]{Cohn+2021}
{Cohn}, H.~N., {Lugger}, P.~M., {Zhao}, Y., {et~al.} 2021, \mnras, 508, 2823,
  \dodoi{10.1093/mnras/stab2636}

\bibitem[{{Correnti} {et~al.}(2016){Correnti}, {Gennaro}, {Kalirai}, {Brown},
  \& {Calamida}}]{Correnti+2016}
{Correnti}, M., {Gennaro}, M., {Kalirai}, J.~S., {Brown}, T.~M., \& {Calamida},
  A. 2016, \apj, 823, 18, \dodoi{10.3847/0004-637X/823/1/18}

\bibitem[{{Dan} {et~al.}(2014){Dan}, {Rosswog}, {Br{\"u}ggen}, \&
  {Podsiadlowski}}]{Dan+2014}
{Dan}, M., {Rosswog}, S., {Br{\"u}ggen}, M., \& {Podsiadlowski}, P. 2014,
  \mnras, 438, 14, \dodoi{10.1093/mnras/stt1766}

\bibitem[{{Davies} \& {Benz}(1995)}]{Davies_Benz_1995}
{Davies}, M.~B., \& {Benz}, W. 1995, \mnras, 276, 876,
  \dodoi{10.1093/mnras/276.3.876}

\bibitem[{{Davies} {et~al.}(1992){Davies}, {Benz}, \& {Hills}}]{Davies+1992}
{Davies}, M.~B., {Benz}, W., \& {Hills}, J.~G. 1992, \apj, 401, 246,
  \dodoi{10.1086/172056}

\bibitem[{{Duquennoy} \& {Mayor}(1991)}]{Duquennoy_Mayor_1991}
{Duquennoy}, A., \& {Mayor}, M. 1991, \aap, 248, 485

\bibitem[{{Farr} {et~al.}(2011){Farr}, {Sravan}, {Cantrell}, {Kreidberg},
  {Bailyn}, {Mandel}, \& {Kalogera}}]{Farr+2011}
{Farr}, W.~M., {Sravan}, N., {Cantrell}, A., {et~al.} 2011, \apj, 741, 103,
  \dodoi{10.1088/0004-637X/741/2/103}

\bibitem[{{Ferrario} {et~al.}(2015){Ferrario}, {de Martino}, \&
  {G{\"a}nsicke}}]{Ferrario2015}
{Ferrario}, L., {de Martino}, D., \& {G{\"a}nsicke}, B.~T. 2015, \ssr, 191,
  111, \dodoi{10.1007/s11214-015-0152-0}

\bibitem[{Fregeau {et~al.}(2004)Fregeau, Cheung, Portegies~Zwart, \&
  Rasio}]{fregeau2004stellar}
Fregeau, J.~M., Cheung, P., Portegies~Zwart, S., \& Rasio, F. 2004, \mnras,
  352, 1

\bibitem[{Fregeau {et~al.}(2003)Fregeau, Gurkan, Joshi, \&
  Rasio}]{Fregeau_2003}
Fregeau, J.~M., Gurkan, M.~A., Joshi, K.~J., \& Rasio, F.~A. 2003, \apj, 593,
  772.
\newblock \url{http://dx.doi.org/10.1086/376593}

\bibitem[{Fregeau \& Rasio(2007)}]{fregeau2007monte}
Fregeau, J.~M., \& Rasio, F.~A. 2007, \apj, 658, 1047

\bibitem[{{Freire} {et~al.}(2008{\natexlab{a}}){Freire}, {Ransom}, {B{\'e}gin},
  {Stairs}, {Hessels}, {Frey}, \& {Camilo}}]{Freire+2008ngc64s}
{Freire}, P. C.~C., {Ransom}, S.~M., {B{\'e}gin}, S., {et~al.}
  2008{\natexlab{a}}, \apj, 675, 670, \dodoi{10.1086/526338}

\bibitem[{{Freire} {et~al.}(2008{\natexlab{b}}){Freire}, {Wolszczan}, {van den
  Berg}, \& {Hessels}}]{Freire+2008M5}
{Freire}, P. C.~C., {Wolszczan}, A., {van den Berg}, M., \& {Hessels}, J. W.~T.
  2008{\natexlab{b}}, \apj, 679, 1433, \dodoi{10.1086/587832}

\bibitem[{{Freire} {et~al.}(2017){Freire}, {Ridolfi}, {Kramer}, {Jordan},
  {Manchester}, {Torne}, {Sarkissian}, {Heinke}, {D'Amico}, {Camilo},
  {Lorimer}, \& {Lyne}}]{Freire+2017}
{Freire}, P.~C.~C., {Ridolfi}, A., {Kramer}, M., {et~al.} 2017, \mnras, 471,
  857, \dodoi{10.1093/mnras/stx1533}

\bibitem[{{Garc{\'\i}a-Berro} {et~al.}(2012){Garc{\'\i}a-Berro},
  {Lor{\'e}n-Aguilar}, {Aznar-Sigu{\'a}n}, {Torres}, {Camacho}, {Althaus},
  {C{\'o}rsico}, {K{\"u}lebi}, \& {Isern}}]{Garcia-Berro+2012}
{Garc{\'\i}a-Berro}, E., {Lor{\'e}n-Aguilar}, P., {Aznar-Sigu{\'a}n}, G.,
  {et~al.} 2012, \apj, 749, 25, \dodoi{10.1088/0004-637X/749/1/25}

\bibitem[{{Gratton} {et~al.}(2003){Gratton}, {Bragaglia}, {Carretta},
  {Clementini}, {Desidera}, {Grundahl}, \& {Lucatello}}]{Gratton+2003}
{Gratton}, R.~G., {Bragaglia}, A., {Carretta}, E., {et~al.} 2003, \aap, 408,
  529, \dodoi{10.1051/0004-6361:20031003}

\bibitem[{{Gratton} {et~al.}(1997){Gratton}, {Fusi Pecci}, {Carretta},
  {Clementini}, {Corsi}, \& {Lattanzi}}]{Gratton+1997}
{Gratton}, R.~G., {Fusi Pecci}, F., {Carretta}, E., {et~al.} 1997, \apj, 491,
  749, \dodoi{10.1086/304987}

\bibitem[{{Hansen} \& {van Horn}(1975)}]{Hansen_vanHorn_1975}
{Hansen}, C.~J., \& {van Horn}, H.~M. 1975, \apj, 195, 735,
  \dodoi{10.1086/153375}

\bibitem[{{Harris}(1996)}]{Harris_1996}
{Harris}, W.~E. 1996, \aj, 112, 1487, \dodoi{10.1086/118116}

\bibitem[{H{\'e}non(1971{\natexlab{a}})}]{henon1971monte}
H{\'e}non, M. 1971{\natexlab{a}}, in International Astronomical Union
  Colloquium, Vol.~10, Cambridge University Press, 151--167

\bibitem[{H{\'e}non(1971{\natexlab{b}})}]{henon1971montecluster}
H{\'e}non, M. 1971{\natexlab{b}}, \apss, 13, 284

\bibitem[{{Hjellming} \& {Webbink}(1987)}]{Hjellming_Webbink_1987}
{Hjellming}, M.~S., \& {Webbink}, R.~F. 1987, \apj, 318, 794,
  \dodoi{10.1086/165412}

\bibitem[{{Hobbs} {et~al.}(2005){Hobbs}, {Lorimer}, {Lyne}, \&
  {Kramer}}]{Hobbs+2005}
{Hobbs}, G., {Lorimer}, D.~R., {Lyne}, A.~G., \& {Kramer}, M. 2005, \mnras,
  360, 974, \dodoi{10.1111/j.1365-2966.2005.09087.x}

\bibitem[{Hurley {et~al.}(2000)Hurley, Pols, \& Tout}]{hurley2000comprehensive}
Hurley, J.~R., Pols, O.~R., \& Tout, C.~A. 2000, \mnras, 315, 543

\bibitem[{Hurley {et~al.}(2002)Hurley, Tout, \& Pols}]{hurley2002evolution}
Hurley, J.~R., Tout, C.~A., \& Pols, O.~R. 2002, \mnras, 329, 897

\bibitem[{{Iben} \& {Tutukov}(1984)}]{Iben_Tutukov_1984}
{Iben}, I., J., \& {Tutukov}, A.~V. 1984, \apjs, 54, 335,
  \dodoi{10.1086/190932}

\bibitem[{{Igoshev} {et~al.}(2021){Igoshev}, {Popov}, \&
  {Hollerbach}}]{Igoshev+2021}
{Igoshev}, A.~P., {Popov}, S.~B., \& {Hollerbach}, R. 2021, Universe, 7, 351,
  \dodoi{10.3390/universe7090351}

\bibitem[{{Jacoby} {et~al.}(2006){Jacoby}, {Cameron}, {Jenet}, {Anderson},
  {Murty}, \& {Kulkarni}}]{Jacoby+2006}
{Jacoby}, B.~A., {Cameron}, P.~B., {Jenet}, F.~A., {et~al.} 2006, \apjl, 644,
  L113, \dodoi{10.1086/505742}

\bibitem[{Joshi {et~al.}(2001)Joshi, Nave, \& Rasio}]{Joshi_2001}
Joshi, K.~J., Nave, C.~P., \& Rasio, F.~A. 2001, \apj, 550, 691.
\newblock \url{http://dx.doi.org/10.1086/319771}

\bibitem[{Joshi {et~al.}(2000)Joshi, Rasio, \& Portegies~Zwart}]{Joshi_2000}
Joshi, K.~J., Rasio, F.~A., \& Portegies~Zwart, S. 2000, \apj, 540, 969.
\newblock \url{http://dx.doi.org/10.1086/309350}

\bibitem[{{Katz} \& {Dong}(2012)}]{KatzDong2012}
{Katz}, B., \& {Dong}, S. 2012, arXiv e-prints, arXiv:1211.4584,
  \dodoi{10.48550/arXiv.1211.4584}

\bibitem[{{Kiel} \& {Hurley}(2009)}]{Kiel_Hurley_2009}
{Kiel}, P.~D., \& {Hurley}, J.~R. 2009, \mnras, 395, 2326,
  \dodoi{10.1111/j.1365-2966.2009.14711.x}

\bibitem[{{Kiel} {et~al.}(2008){Kiel}, {Hurley}, {Bailes}, \&
  {Murray}}]{Kiel+2008}
{Kiel}, P.~D., {Hurley}, J.~R., {Bailes}, M., \& {Murray}, J.~R. 2008, \mnras,
  388, 393, \dodoi{10.1111/j.1365-2966.2008.13402.x}

\bibitem[{{King}(1966)}]{King1966}
{King}, I.~R. 1966, \aj, 71, 64, \dodoi{10.1086/109857}

\bibitem[{{Kirsten} {et~al.}(2022){Kirsten}, {Marcote}, {Nimmo}, {Hessels},
  {Bhardwaj}, {Tendulkar}, {Keimpema}, {Yang}, {Snelders}, {Scholz},
  {Pearlman}, {Law}, {Peters}, {Giroletti}, {Paragi}, {Bassa}, {Hewitt},
  {Bach}, {Bezrukovs}, {Burgay}, {Buttaccio}, {Conway}, {Corongiu}, {Feiler},
  {Forss{\'e}n}, {Gawro{\'n}ski}, {Karuppusamy}, {Kharinov}, {Lindqvist},
  {Maccaferri}, {Melnikov}, {Ould-Boukattine}, {Possenti}, {Surcis}, {Wang},
  {Yuan}, {Aggarwal}, {Anna-Thomas}, {Bower}, {Blaauw}, {Burke-Spolaor},
  {Cassanelli}, {Clarke}, {Fonseca}, {Gaensler}, {Gopinath}, {Kaspi}, {Kassim},
  {Lazio}, {Leung}, {Li}, {Lin}, {Masui}, {Mckinven}, {Michilli}, {Mikhailov},
  {Ng}, {Orbidans}, {Pen}, {Petroff}, {Rahman}, {Ransom}, {Shin}, {Smith},
  {Stairs}, \& {Vlemmings}}]{Kirsten+2022}
{Kirsten}, F., {Marcote}, B., {Nimmo}, K., {et~al.} 2022, \nat, 602, 585,
  \dodoi{10.1038/s41586-021-04354-w}

\bibitem[{{Kremer} {et~al.}(2023{\natexlab{a}}){Kremer}, {Fuller}, {Piro}, \&
  {Ransom}}]{Kremer+2023}
{Kremer}, K., {Fuller}, J., {Piro}, A.~L., \& {Ransom}, S.~M.
  2023{\natexlab{a}}, arXiv e-prints, arXiv:2305.11933,
  \dodoi{10.48550/arXiv.2305.11933}

\bibitem[{{Kremer} {et~al.}(2023{\natexlab{b}}){Kremer}, {Li}, {Lu}, {Piro}, \&
  {Zhang}}]{Kremer+2023_detectfrb}
{Kremer}, K., {Li}, D., {Lu}, W., {Piro}, A.~L., \& {Zhang}, B.
  2023{\natexlab{b}}, \apj, 944, 6, \dodoi{10.3847/1538-4357/acabbf}

\bibitem[{{Kremer} {et~al.}(2019){Kremer}, {Lu}, {Rodriguez}, {Lachat}, \&
  {Rasio}}]{Kremer+2019tde}
{Kremer}, K., {Lu}, W., {Rodriguez}, C.~L., {Lachat}, M., \& {Rasio}, F.~A.
  2019, \apj, 881, 75, \dodoi{10.3847/1538-4357/ab2e0c}

\bibitem[{{Kremer} {et~al.}(2021{\natexlab{a}}){Kremer}, {Piro}, \&
  {Li}}]{Kremer+2021frb}
{Kremer}, K., {Piro}, A.~L., \& {Li}, D. 2021{\natexlab{a}}, \apjl, 917, L11,
  \dodoi{10.3847/2041-8213/ac13a0}

\bibitem[{{Kremer} {et~al.}(2021{\natexlab{b}}){Kremer}, {Rui}, {Weatherford},
  {Chatterjee}, {Fragione}, {Rasio}, {Rodriguez}, \& {Ye}}]{Kremer+2021WD}
{Kremer}, K., {Rui}, N.~Z., {Weatherford}, N.~C., {et~al.} 2021{\natexlab{b}},
  \apj, 917, 28, \dodoi{10.3847/1538-4357/ac06d4}

\bibitem[{{Kremer} {et~al.}(2020{\natexlab{a}}){Kremer}, {Ye}, {Chatterjee},
  {Rodriguez}, \& {Rasio}}]{Kremer+2020bhburning}
{Kremer}, K., {Ye}, C.~S., {Chatterjee}, S., {Rodriguez}, C.~L., \& {Rasio},
  F.~A. 2020{\natexlab{a}}, in Star Clusters: From the Milky Way to the Early
  Universe, ed. A.~{Bragaglia}, M.~{Davies}, A.~{Sills}, \& E.~{Vesperini},
  Vol. 351, 357--366, \dodoi{10.1017/S1743921319007269}

\bibitem[{{Kremer} {et~al.}(2022){Kremer}, {Ye}, {K{\i}ro{\u{g}}lu},
  {Lombardi}, {Ransom}, \& {Rasio}}]{Kremer_nstde_2022}
{Kremer}, K., {Ye}, C.~S., {K{\i}ro{\u{g}}lu}, F., {et~al.} 2022, arXiv
  e-prints, arXiv:2204.07169.
\newblock \doarXiv{2204.07169}

\bibitem[{{Kremer} {et~al.}(2020{\natexlab{b}}){Kremer}, {Ye}, {Rui},
  {Weatherford}, {Chatterjee}, {Fragione}, {Rodriguez}, {Spera}, \&
  {Rasio}}]{Kremer+2020catalog}
{Kremer}, K., {Ye}, C.~S., {Rui}, N.~Z., {et~al.} 2020{\natexlab{b}}, \apjs,
  247, 48, \dodoi{10.3847/1538-4365/ab7919}

\bibitem[{Kroupa(2001)}]{Kroupa2001}
Kroupa, P. 2001, \mnras, 322, 231

\bibitem[{{Kumar} {et~al.}(2008){Kumar}, {Narayan}, \& {Johnson}}]{Kumar2008}
{Kumar}, P., {Narayan}, R., \& {Johnson}, J.~L. 2008, \mnras, 388, 1729,
  \dodoi{10.1111/j.1365-2966.2008.13493.x}

\bibitem[{{Lam} {et~al.}(2022){Lam}, {Lu}, {Udalski}, {Bond}, {Bennett},
  {Skowron}, {Mr{\'o}z}, {Poleski}, {Sumi}, {Szyma{\'n}ski}, {Koz{\l}owski},
  {Pietrukowicz}, {Soszy{\'n}ski}, {Ulaczyk}, {Wyrzykowski}, {Miyazaki},
  {Suzuki}, {Koshimoto}, {Rattenbury}, {Hosek}, {Abe}, {Barry}, {Bhattacharya},
  {Fukui}, {Fujii}, {Hirao}, {Itow}, {Kirikawa}, {Kondo}, {Matsubara},
  {Matsumoto}, {Muraki}, {Olmschenk}, {Ranc}, {Okamura}, {Satoh}, {Silva},
  {Toda}, {Tristram}, {Vandorou}, {Yama}, {Abrams}, {Agarwal}, {Rose}, \&
  {Terry}}]{Lam+2022}
{Lam}, C.~Y., {Lu}, J.~R., {Udalski}, A., {et~al.} 2022, \apjl, 933, L23,
  \dodoi{10.3847/2041-8213/ac7442}

\bibitem[{{Lee} {et~al.}(1996){Lee}, {Kim}, \& {Kang}}]{Lee+1996}
{Lee}, H.~M., {Kim}, S.~S., \& {Kang}, H. 1996, Journal of Korean Astronomical
  Society, 29, 19, \dodoi{10.48550/arXiv.astro-ph/9603137}

\bibitem[{{Leigh} {et~al.}(2023){Leigh}, {Ye}, {Grondin}, {Fragione}, {Webb},
  \& {Heinke}}]{Leigh+2023}
{Leigh}, N. W.~C., {Ye}, C.~S., {Grondin}, S.~M., {et~al.} 2023, arXiv
  e-prints, arXiv:2309.13122, \dodoi{10.48550/arXiv.2309.13122}

\bibitem[{{Levan} {et~al.}(2006){Levan}, {Wynn}, {Chapman}, {Davies}, {King},
  {Priddey}, \& {Tanvir}}]{Levan+2006}
{Levan}, A.~J., {Wynn}, G.~A., {Chapman}, R., {et~al.} 2006, \mnras, 368, L1,
  \dodoi{10.1111/j.1745-3933.2006.00144.x}

\bibitem[{{Li} \& {Pen}(2023)}]{Li_Pen_2023}
{Li}, D., \& {Pen}, U.-L. 2023, arXiv e-prints, arXiv:2309.06328,
  \dodoi{10.48550/arXiv.2309.06328}

\bibitem[{{Libralato} {et~al.}(2022){Libralato}, {Bellini}, {Vesperini},
  {Piotto}, {Milone}, {van der Marel}, {Anderson}, {Aparicio}, {Barbuy},
  {Bedin}, {Borsato}, {Cassisi}, {Dalessandro}, {Ferraro}, {King}, {Lanzoni},
  {Nardiello}, {Ortolani}, {Sarajedini}, \& {Sohn}}]{Libralato+2022}
{Libralato}, M., {Bellini}, A., {Vesperini}, E., {et~al.} 2022, \apj, 934, 150,
  \dodoi{10.3847/1538-4357/ac7727}

\bibitem[{{Lorimer}(2008)}]{Lorimer_2008}
{Lorimer}, D.~R. 2008, Living Reviews in Relativity, 11, 8,
  \dodoi{10.12942/lrr-2008-8}

\bibitem[{{Lu} {et~al.}(2022){Lu}, {Beniamini}, \& {Kumar}}]{Lu2022}
{Lu}, W., {Beniamini}, P., \& {Kumar}, P. 2022, \mnras, 510, 1867,
  \dodoi{10.1093/mnras/stab3500}

\bibitem[{{Lynch} {et~al.}(2012){Lynch}, {Freire}, {Ransom}, \&
  {Jacoby}}]{Lynch+2012}
{Lynch}, R.~S., {Freire}, P. C.~C., {Ransom}, S.~M., \& {Jacoby}, B.~A. 2012,
  \apj, 745, 109, \dodoi{10.1088/0004-637X/745/2/109}

\bibitem[{{Manchester} {et~al.}(2005){Manchester}, {Hobbs}, {Teoh}, \&
  {Hobbs}}]{Manchester+2005}
{Manchester}, R.~N., {Hobbs}, G.~B., {Teoh}, A., \& {Hobbs}, M. 2005, \aj, 129,
  1993, \dodoi{10.1086/428488}

\bibitem[{{Maoz} \& {Mannucci}(2012)}]{Maoz_Mannucci_2012}
{Maoz}, D., \& {Mannucci}, F. 2012, \pasa, 29, 447, \dodoi{10.1071/AS11052}

\bibitem[{{Maoz} {et~al.}(2014){Maoz}, {Mannucci}, \& {Nelemans}}]{Maoz+2014}
{Maoz}, D., {Mannucci}, F., \& {Nelemans}, G. 2014, \araa, 52, 107,
  \dodoi{10.1146/annurev-astro-082812-141031}

\bibitem[{{Margalit} {et~al.}(2019){Margalit}, {Berger}, \&
  {Metzger}}]{Margalit2019}
{Margalit}, B., {Berger}, E., \& {Metzger}, B.~D. 2019, \apj, 886, 110,
  \dodoi{10.3847/1538-4357/ab4c31}

\bibitem[{{Marsh} {et~al.}(2004){Marsh}, {Nelemans}, \& {Steeghs}}]{Marsh2004}
{Marsh}, T.~R., {Nelemans}, G., \& {Steeghs}, D. 2004, \mnras, 350, 113,
  \dodoi{10.1111/j.1365-2966.2004.07564.x}

\bibitem[{{Metzger} {et~al.}(2008){Metzger}, {Piro}, \&
  {Quataert}}]{Metzger2008}
{Metzger}, B.~D., {Piro}, A.~L., \& {Quataert}, E. 2008, \mnras, 390, 781,
  \dodoi{10.1111/j.1365-2966.2008.13789.x}

\bibitem[{{Morscher} {et~al.}(2015){Morscher}, {Pattabiraman}, {Rodriguez},
  {Rasio}, \& {Umbreit}}]{Morscher+2015}
{Morscher}, M., {Pattabiraman}, B., {Rodriguez}, C., {Rasio}, F.~A., \&
  {Umbreit}, S. 2015, \apj, 800, 9, \dodoi{10.1088/0004-637X/800/1/9}

\bibitem[{{Neijssel} {et~al.}(2019){Neijssel}, {Vigna-G{\'o}mez}, {Stevenson},
  {Barrett}, {Gaebel}, {Broekgaarden}, {de Mink}, {Sz{\'e}csi}, {Vinciguerra},
  \& {Mandel}}]{Neijssel+2019}
{Neijssel}, C.~J., {Vigna-G{\'o}mez}, A., {Stevenson}, S., {et~al.} 2019,
  \mnras, 490, 3740, \dodoi{10.1093/mnras/stz2840}

\bibitem[{{Nelemans} {et~al.}(2001){Nelemans}, {Yungelson}, \& {Portegies
  Zwart}}]{Nelemans2001}
{Nelemans}, G., {Yungelson}, L.~R., \& {Portegies Zwart}, S.~F. 2001, \aap,
  375, 890, \dodoi{10.1051/0004-6361:20010683}

\bibitem[{{Nomoto}(1984)}]{nomoto1984evolution}
{Nomoto}, K. 1984, \apj, 277, 791, \dodoi{10.1086/161749}

\bibitem[{{Nomoto}(1987)}]{nomoto1987evolution}
---. 1987, \apj, 322, 206, \dodoi{10.1086/165716}

\bibitem[{{Nomoto} \& {Iben}(1985)}]{Nomoto_Iben_1985}
{Nomoto}, K., \& {Iben}, I., J. 1985, \apj, 297, 531, \dodoi{10.1086/163547}

\bibitem[{Nomoto \& Kondo(1991)}]{nomoto1991conditions}
Nomoto, K., \& Kondo, Y. 1991, \apj, 367, L19

\bibitem[{{{\"O}zel} {et~al.}(2010){{\"O}zel}, {Psaltis}, {Narayan}, \&
  {McClintock}}]{Ozel+2010}
{{\"O}zel}, F., {Psaltis}, D., {Narayan}, R., \& {McClintock}, J.~E. 2010,
  \apj, 725, 1918, \dodoi{10.1088/0004-637X/725/2/1918}

\bibitem[{{Pakmor} {et~al.}(2012){Pakmor}, {Kromer}, {Taubenberger}, {Sim},
  {R{\"o}pke}, \& {Hillebrandt}}]{Pakmor+2012}
{Pakmor}, R., {Kromer}, M., {Taubenberger}, S., {et~al.} 2012, \apjl, 747, L10,
  \dodoi{10.1088/2041-8205/747/1/L10}

\bibitem[{{Papish} {et~al.}(2015){Papish}, {Soker}, \& {Bukay}}]{Papish+2015}
{Papish}, O., {Soker}, N., \& {Bukay}, I. 2015, \mnras, 449, 288,
  \dodoi{10.1093/mnras/stv345}

\bibitem[{{Ransom} {et~al.}(2005){Ransom}, {Hessels}, {Stairs}, {Freire},
  {Camilo}, {Kaspi}, \& {Kaplan}}]{Ransom+2005}
{Ransom}, S.~M., {Hessels}, J. W.~T., {Stairs}, I.~H., {et~al.} 2005, Science,
  307, 892, \dodoi{10.1126/science.1108632}

\bibitem[{{Ridolfi} {et~al.}(2019){Ridolfi}, {Freire}, {Gupta}, \&
  {Ransom}}]{Ridolfi+2019}
{Ridolfi}, A., {Freire}, P.~C.~C., {Gupta}, Y., \& {Ransom}, S.~M. 2019,
  \mnras, 490, 3860, \dodoi{10.1093/mnras/stz2645}

\bibitem[{{Ridolfi} {et~al.}(2021){Ridolfi}, {Gautam}, {Freire}, {Ransom},
  {Buchner}, {Possenti}, {Venkatraman Krishnan}, {Bailes}, {Kramer},
  {Stappers}, {Abbate}, {Barr}, {Burgay}, {Camilo}, {Corongiu}, {Jameson},
  {Padmanabh}, {Vleeschower}, {Champion}, {Chen}, {Geyer}, {Karastergiou},
  {Karuppusamy}, {Parthasarathy}, {Reardon}, {Serylak}, {Shannon}, \&
  {Spiewak}}]{Ridolfi+2021}
{Ridolfi}, A., {Gautam}, T., {Freire}, P.~C.~C., {et~al.} 2021, \mnras, 504,
  1407, \dodoi{10.1093/mnras/stab790}

\bibitem[{Rodriguez {et~al.}(2018b)Rodriguez, Amaro-Seoane, Chatterjee, Kremer,
  Rasio, Samsing, Claire, \& Zevin}]{rodriguez2018postb}
Rodriguez, C.~L., Amaro-Seoane, P., Chatterjee, S., {et~al.} 2018b, \prd, 98,
  123005

\bibitem[{{Rodriguez} {et~al.}(2018a){Rodriguez}, {Amaro-Seoane}, {Chatterjee},
  \& {Rasio}}]{Rodriguez+repeated2018}
{Rodriguez}, C.~L., {Amaro-Seoane}, P., {Chatterjee}, S., \& {Rasio}, F.~A.
  2018a, \prl, 120, 151101, \dodoi{10.1103/PhysRevLett.120.151101}

\bibitem[{{Rodriguez} {et~al.}(2016){Rodriguez}, {Morscher}, {Wang},
  {Chatterjee}, {Rasio}, \& {Spurzem}}]{Rodriguez+2016million}
{Rodriguez}, C.~L., {Morscher}, M., {Wang}, L., {et~al.} 2016, \mnras, 463,
  2109, \dodoi{10.1093/mnras/stw2121}

\bibitem[{{Rodriguez} {et~al.}(2022){Rodriguez}, {Weatherford}, {Coughlin},
  {Amaro-Seoane}, {Breivik}, {Chatterjee}, {Fragione}, {K{\i}ro{\u{g}}lu},
  {Kremer}, {Rui}, {Ye}, {Zevin}, \& {Rasio}}]{CMC1}
{Rodriguez}, C.~L., {Weatherford}, N.~C., {Coughlin}, S.~C., {et~al.} 2022,
  \apjs, 258, 22, \dodoi{10.3847/1538-4365/ac2edf}

\bibitem[{{Saio} \& {Nomoto}(1985)}]{Saio_Nomoto_1985}
{Saio}, H., \& {Nomoto}, K. 1985, \aap, 150, L21

\bibitem[{{Saio} \& {Nomoto}(1998)}]{Saio_Nomoto_1998}
---. 1998, \apj, 500, 388, \dodoi{10.1086/305696}

\bibitem[{{Saio} \& {Nomoto}(2004)}]{Saio_Nomoto_2004}
---. 2004, \apj, 615, 444, \dodoi{10.1086/423976}

\bibitem[{{Schwab}(2021)}]{Schwab_2021}
{Schwab}, J. 2021, \apj, 906, 53, \dodoi{10.3847/1538-4357/abc87e}

\bibitem[{{Schwab} {et~al.}(2015){Schwab}, {Quataert}, \&
  {Bildsten}}]{Schwab+2015}
{Schwab}, J., {Quataert}, E., \& {Bildsten}, L. 2015, \mnras, 453, 1910,
  \dodoi{10.1093/mnras/stv1804}

\bibitem[{{Schwab} {et~al.}(2016){Schwab}, {Quataert}, \&
  {Kasen}}]{Schwab+2016}
{Schwab}, J., {Quataert}, E., \& {Kasen}, D. 2016, \mnras, 463, 3461,
  \dodoi{10.1093/mnras/stw2249}

\bibitem[{{Schwab} {et~al.}(2012){Schwab}, {Shen}, {Quataert}, {Dan}, \&
  {Rosswog}}]{Schwab2012}
{Schwab}, J., {Shen}, K.~J., {Quataert}, E., {Dan}, M., \& {Rosswog}, S. 2012,
  \mnras, 427, 190, \dodoi{10.1111/j.1365-2966.2012.21993.x}

\bibitem[{{Shapiro} \& {Teukolsky}(1983)}]{Shapiro_Teukolsky_1983}
{Shapiro}, S.~L., \& {Teukolsky}, S.~A. 1983, {Black holes, white dwarfs, and
  neutron stars : the physics of compact objects}

\bibitem[{{Shen} {et~al.}(2012){Shen}, {Bildsten}, {Kasen}, \&
  {Quataert}}]{Shen+2012}
{Shen}, K.~J., {Bildsten}, L., {Kasen}, D., \& {Quataert}, E. 2012, \apj, 748,
  35, \dodoi{10.1088/0004-637X/748/1/35}

\bibitem[{{Souza} {et~al.}(2020){Souza}, {Kerber}, {Barbuy},
  {P{\'e}rez-Villegas}, {Oliveira}, \& {Nardiello}}]{Souza+2020}
{Souza}, S.~O., {Kerber}, L.~O., {Barbuy}, B., {et~al.} 2020, \apj, 890, 38,
  \dodoi{10.3847/1538-4357/ab6a0f}

\bibitem[{{Taam}(1985)}]{Taam1985}
{Taam}, R.~E. 1985, Annual Review of Nuclear and Particle Science, 35, 1,
  \dodoi{10.1146/annurev.ns.35.120185.000245}

\bibitem[{{Thompson} {et~al.}(2019){Thompson}, {Kochanek}, {Stanek}, {Badenes},
  {Post}, {Jayasinghe}, {Latham}, {Bieryla}, {Esquerdo}, {Berlind}, {Calkins},
  {Tayar}, {Lindegren}, {Johnson}, {Holoien}, {Auchettl}, \&
  {Covey}}]{Thompson+2019}
{Thompson}, T.~A., {Kochanek}, C.~S., {Stanek}, K.~Z., {et~al.} 2019, Science,
  366, 637, \dodoi{10.1126/science.aau4005}

\bibitem[{{Thorsett} \& {Chakrabarty}(1999)}]{Thorsett_Chakrabarty_1999}
{Thorsett}, S.~E., \& {Chakrabarty}, D. 1999, \apj, 512, 288,
  \dodoi{10.1086/306742}

\bibitem[{{Trager} {et~al.}(1995){Trager}, {King}, \&
  {Djorgovski}}]{Trager1995}
{Trager}, S.~C., {King}, I.~R., \& {Djorgovski}, S. 1995, \aj, 109, 218,
  \dodoi{10.1086/117268}

\bibitem[{Umbreit {et~al.}(2012)Umbreit, Fregeau, Chatterjee, \&
  Rasio}]{Umbreit_2012}
Umbreit, S., Fregeau, J.~M., Chatterjee, S., \& Rasio, F.~A. 2012, \apj, 750,
  31.
\newblock \url{http://dx.doi.org/10.1088/0004-637X/750/1/31}

\bibitem[{{Usov}(1992)}]{Usov_1992}
{Usov}, V.~V. 1992, \nat, 357, 472, \dodoi{10.1038/357472a0}

\bibitem[{{Vasiliev} \& {Baumgardt}(2021)}]{Vasiliev+2021}
{Vasiliev}, E., \& {Baumgardt}, H. 2021, \mnras, 505, 5978,
  \dodoi{10.1093/mnras/stab1475}

\bibitem[{{Verbunt} \& {Freire}(2014)}]{Verbunt_Freire_2014}
{Verbunt}, F., \& {Freire}, P. C.~C. 2014, \aap, 561, A11,
  \dodoi{10.1051/0004-6361/201321177}

\bibitem[{{Webbink}(1984)}]{Webbink_1984}
{Webbink}, R.~F. 1984, \apj, 277, 355, \dodoi{10.1086/161701}

\bibitem[{{Ye} {et~al.}(2019){Ye}, {Kremer}, {Chatterjee}, {Rodriguez}, \&
  {Rasio}}]{Ye_msp_2019}
{Ye}, C.~S., {Kremer}, K., {Chatterjee}, S., {Rodriguez}, C.~L., \& {Rasio},
  F.~A. 2019, \apj, 877, 122, \dodoi{10.3847/1538-4357/ab1b21}

\bibitem[{{Ye} {et~al.}(2022){Ye}, {Kremer}, {Rodriguez}, {Rui}, {Weatherford},
  {Chatterjee}, {Rasio}, \& {Fragione}}]{Ye_47tuc_2021}
{Ye}, C.~S., {Kremer}, K., {Rodriguez}, C.~L., {et~al.} 2022, \apj, 931, 84,
  \dodoi{10.3847/1538-4357/ac5b0b}

\end{thebibliography}
\bibliographystyle{aasjournal}

\end{document}